\pdfoutput=1
\RequirePackage{fixltx2e}[2005/12/01] 
\documentclass[useAMS,usenatbib]{mn2e}
\usepackage{amsmath}
\usepackage[varg]{txfonts}
\usepackage{astrojournals}
\usepackage{graphicx}

\usepackage{microtype}

\usepackage{xcolor}
\definecolor{Red}{rgb}{0.7,0.1,0.1}
\DeclareRobustCommand\New[1]{#1}
\DeclareRobustCommand\Newest[1]{#1}
\DeclareRobustCommand\Old[1]{}

\newcommand\pcc{\ensuremath{\mathrm{cm}^{-3}}}
\newcommand\kms{\ensuremath{\mathrm{km\ s^{-1}}}}
\newcommand\heatunits{\ensuremath{\mathrm{erg\ s^{-1}\ cm^{-3}}}}
\newcounter{ionstage}
\newcommand{\ion}[2]{\setcounter{ionstage}{#2}% 
  \ensuremath{\mathrm{#1\,\scriptstyle\Roman{ionstage}}}}

\begin{document}

%% Title Page, etc,

\title[Photoionisation of magnetised globules]
{Radiation-magnetohydrodynamic simulations of the photoionisation of magnetised globules\thanks{Based in part on numerical simulations carried out using the Kan~Balam supercomputer, operated by the Departamento de Superc\'omputo, Direcci\'on General de Servicios de C\'omputo Acad\'emico, Universidad Nacional Aut\'onoma de M\'exico.}}

\author[W.~Henney et al.\@] {%
  William J. Henney,\textsuperscript{1} S. Jane
  Arthur,\textsuperscript{1} Fabio De Colle,\textsuperscript{2,3} and
  Garrelt Mellema\textsuperscript{4}
  \\
  \textsuperscript{1}%
  Centro de Radioastronom\'{\i}a y Astrof\'{\i}sica, Universidad
  Nacional Aut\'onoma de M\'exico, Apartado Postal 3-72, 58090
  Morelia, Michoac\'an, M\'exico\\
  \textsuperscript{2}%
  School of Cosmic Physics, Dublin Institute for Advanced Studies, 31
  Fitzwilliam Place, Dublin, Ireland\\
  \textsuperscript{3}%
  Department of Astronomy and Astrophysics, University of California Santa
  Cruz, Santa Cruz, CA 95064, USA\\
  \textsuperscript{4}%
  Department of Astronomy, Stockholm University, SE-106 91 Stockholm,
  Sweden
}

% \email{w.henney,j.arthur@astrosmo.unam.mx}

\maketitle

\begin{abstract}
  We present the first three-dimensional radiation-magnetohydrodynamic simulations of the photoionisation of a dense, magnetised molecular globule by an external source of ultraviolet radiation. We find that, for the case of a strong ionising field, significant deviations from the non-magnetic evolution are seen when the initial magnetic field threading the globule has an associated magnetic pressure that is greater than one hundred times the gas pressure. In such a strong-field case, the photoevaporating globule will adopt a flattened or ``curled up'' shape, depending on the initial field orientation, and magnetic confinement of the ionised photoevaporation flow can lead to recombination and subsequent fragmentation during advanced stages of the globule evolution. We find suggestive evidence that such magnetic effects may be important in the formation of bright, bar-like emission features in \ion{H}{2} regions. \New{We include simple but realistic fits to heating and cooling rates in the neutral and molecular gas in the vicinity of a high-mass star cluster and show that the frequently used isothermal approximation can lead to an overestimate of the importance of gravitational instability in the radiatively imploded globule. For globules within 2 parsecs of a high-mass star cluster, we find that heating by stellar x rays prevents the molecular gas from cooling below 50~K.} 
\end{abstract}


%% Text of paper

\section{Introduction}
\label{sec:introduction}

When ionising radiation propagates into a non-uniform medium, the resultant structure and evolution is very different from the textbook development of a spherical \ion{H}{2} region \citep{1939ApJ....89..526S, 1997pism.book.....D}. Dense condensations in the neutral gas will slow down the propagation of the ionisation front in some directions, causing it to ``wrap around'' the obstacle (see \citealp{2007dmsf.book..103H}, \S~2.4, for a summary). A simple idealisation of this phenomenon is the model of a photoevaporating globule: a dense, isolated cloud of neutral/molecular gas that is illuminated from one side by a source of ionising radiation.  The earliest analytical studies of globule photoevaporation \citep{1955ApJ...121....6O, 1968Ap&SS...1..388D, 1969Phy....41..172K} already identified the three principal aspects of their evolution: (1) the radiation-driven implosion of the neutral/molecular gas, possibly giving rise to enhanced star formation; (2) the transonic flow of ionised gas back towards the radiation source, capable of driving turbulence in the \ion{H}{2} region; (3) the acceleration of the residual neutral globule away from the radiation source due to the back reaction of the ionised photoevaporation flow (rocket effect). Later semi-analytical and numerical work \citep{1982ApJ...260..183S, 1981A&A....99..305T, 1989ApJ...346..735B,1990ApJ...354..529B} investigated the globule evolution in greater detail and exhaustively explored the two-dimensional parameter space of initial globule column density versus ionisation parameter (ratio of ionising photon density to particle density). More recent studies have concentrated on the effects of the diffuse radiation field and multiple ionising sources \citep{1998ApJ...502..695C, 2001A&A...369..263P, 2006RMxAA..42..203C}, together with detailed studies of the evolution of non-uniform globules, applied to gravitational collapse and star formation \citep{2003MNRAS.338..545K, 2005RMxAA..41..443G, 2007MNRAS.377..383E, 2007A&A...467..657M}.

The principal application of globule models has been to the study of structures such as bright rims, pillars, and ``elephant trunks'', which are frequently found at the periphery of \ion{H}{2} regions around young massive stars \citep{1956BAN....13...77P,  1984A&A...135...81B, 1998A&A...332L...5C, 2001MNRAS.327..788W}. Two principal modes of formation have been considered for the globules: they may simply be pre-existing dense cores in the molecular cloud \citep{1983A&A...117..183R, 2006ApJ...647..397M}, or they may be the result of hydrodynamical instabilities of the ionisation front and preceding shocked neutral shell \citep{1954ApJ...120....1S, 1964ApJ...140..112A, 1979ApJ...233..280G, 1996ApJ...469..171G, 2002MNRAS.331..693W, 2006ApJ...647.1151M, 2008ApJ...672..287W}. The equilibrium globule structure is found to be almost identical in the two cases \citep{2001MNRAS.327..788W}. A second domain of application for globule models is to the dense knots seen in some planetary nebulae \citep{1998A&A...331..335M, 2001ApJ...548..288L,  2002AJ....123.3329O, 2005AJ....130..172O}.

Although some previous studies have considered the effects of magnetic fields on globule evolution, most such work has been qualitative and heuristic only \citep{1989ApJ...346..735B, 2003A&A...403..399C, 2005Ap&SS.298..183R}. The only published numerical calculations have been carried out in one or two dimensions \citep{2001MNRAS.325..293W, 2007Ap&SS.307..179W}, which is not sufficient to capture the magnetic field geometry in the general case. One-dimensional studies of magnetohydrodynamic (MHD) ionisation fronts \citep{1998A&A...331.1099R, 2000MNRAS.314..315W, 2001MNRAS.325..293W} have shown that their jump conditions and internal structure can be considerably modified from the pure hydrodynamic case, while a recent three-dimensional simulation of the evolution of an MHD \ion{H}{2} region in a uniform medium \citep{2007ApJ...671..518K} has shown that magnetic fields dominate the dynamics at late times.

In this paper, we present the first three-dimensional radiation-MHD simulations of the evolution of a photoionised globule. In \S~\ref{sec:numerical} we describe the numerical algorithms and initial conditions that we employ. In \S~\ref{sec:results} we describe in detail the results of the simulations with different magnetic field strengths and orientations. In \S~\ref{sec:discussion} we compare our simulations with observations of photoevaporated globules and similar objects.

\section{Numerical simulations}
\label{sec:numerical}

We carry out our simulations using a new code Phab-C\textsuperscript{2}, which couples an Eulerian Godunov MHD code \citep{2005MNRAS.359..164D, 2006A&A...449.1061D} with the radiative transfer/photoionisation code C\textsuperscript{2}-ray \citep{2006NewA...11..374M, 2006ApJ...647..397M}. 

\subsection{Basic equations}
\label{sec:basic-equations}

The equations solved are schematically as follows:
\begin{gather}
 {\partial \rho \over\partial t}
  +\nabla\cdot\left(\,\rho \vec v\,\right)  = 0  \label{mhd1}
  \\
 {\partial \rho \vec v \over\partial t}
  +\nabla\cdot
  \left(\rho \vec v \vec v + p_{\mathrm{tot}} I-\vec B \vec B\right)=0
 \label{mhd2}
 \\
 {\partial e \over\partial t}
  +\nabla\cdot
  \left( \left(e+p_{\mathrm{tot}}\right)\vec v
    - \bigl(\vec v \cdot \vec B\,\bigr)\vec B\right) = \New{H} - L
 \label{mhd3}
 \\
 {\partial \vec B \over\partial t}
  +\nabla\cdot
  \left(\vec v \vec B - \vec B \vec v\right) =0
 \label{mhd4}
\end{gather}
where $\rho$ is the mass density, $\vec v$ is the velocity vector, $p_{\mathrm{tot}} = p_{\mathrm{gas}}+B^2/2$ is the (\(\textrm{magnetic} + \textrm{thermal}\)) total pressure, $I$ is the identity matrix, $\vec B$ is the magnetic field (in units of \(\mathrm{Gauss} / \surd 4\pi \)), $e$ is the total energy defined as $e= \frac{1}{\gamma-1} p_{\mathrm{gas}}+ \frac{1}{2} \rho v^2+ \frac{1}{2} B^2$ (with $\gamma=5/3$), and \(L\) and \New{\(H\)} are respectively the microphysical cooling and heating rates, which are functions of the local gas and radiation conditions.
The above equations represent the conservation of mass (\ref{mhd1}), momentum (\ref{mhd2}), energy (\ref{mhd3}) and magnetic flux (\ref{mhd4}).
They are combined with an equation for hydrogen ionisation/recombination:
\newcommand\Hp{_\mathrm{p}}
\newcommand\Hz{_\mathrm{n}}
\begin{multline}
  {\partial\, n\Hz \over\partial t}
  + \nabla \cdot \left(\,n\Hz \vec v\,\right) = \\
  n\Hp\, n_{\mathrm{e}}\, \alpha(T) - n\Hz \left( n_{\mathrm{e}} C(T)  + 
   \int_{\nu_0}^\infty \!\!\!\sigma_\nu (4 \pi J_\nu / h\nu) \,d\nu \right),  
 \label{mhd5}
\end{multline}
where \(n\Hp\), \(n\Hz\), and \(n_{\mathrm{e}}\) are the number densities of ionised and neutral hydrogen, and electrons, respectively. Additionally, \(\alpha(T)\) and \(C(T)\) are respectively the radiative recombination and collisional ionisation coefficients, while \(\sigma_\nu\) is the photoionization cross-section and \(J_\nu\) is the local mean intensity of the ionising radiation field, both functions of the photon frequency \(\nu\). The direct contribution of a single, point-like radiation source, of luminosity \(\mathcal{L}^*_\nu\) and located at \(\vec r_*\), to the local radiation field at a point \(\vec r\) is given by 
\begin{gather}
  4\pi J^*_\nu(\vec r) = 
  \frac{ \mathcal{L}^*_\nu \, e^{-\tau_\nu} }{ 4 \pi |\vec r - \vec r_*|^2 }, 
  \label{eq:rad}\\
  \intertext{with}
  \tau_\nu = 
  \int_0^{|\vec r - \vec r_*|} \!\!\!  n\Hz (\vec r_* + s \vec e_r) \, \sigma_\nu \, ds ,
  \label{eq:tau}
\end{gather}
where \(\vec e_r\) is the unit vector \((\vec r - \vec r_*)/|\vec r - \vec r_*| \) and \(s\) is the distance along the straight-line path between \(\vec r_*\) and \(\vec r\). The diffuse field due to ground-state recombinations is treated in the standard on-the-spot approximation \citep{2006agna.book.....O}, in which it is not explicitly included in \(J_\nu\) and the case-B value for \(\alpha(T)\) is used.

\subsection{Implementation}
\label{sec:implementation}

All calculations are performed on a fixed, regular Cartesian grid in two or three dimensions. The MHD Riemann solver uses a standard second-order Runge-Kutta method for the time integration and a spatially second-order reconstruction of the primitive variables at the interfaces (except in shocks). Lapidus viscosity is applied to fluxes at cell interfaces, following \citet{1984JCoPh..54..174C}. The constrained transport (CT) method \citep[e.g.,][]{2000JCoPh.161..605T} is used to conserve \(\vec{\nabla} \cdot \vec{B} = 0\) to machine accuracy. 

The C\textsuperscript{2}-ray (Conservative-Causal ray, \citealp{2006NewA...11..374M}) algorithm uses a short-characteristic method to calculate the radiative transfer of ionising radiation (equation~(\ref{eq:tau})), together with an explicitly photon-conserving iterative technique for solving the ionisation rate equation (equation~(\ref{mhd5})), which allows one to use timesteps that are much longer than the ionisation/recombination timescales without loss of accuracy \citep{2006MNRAS.371.1057I}. 

The radiative heating and cooling terms (\New{\(H\)} and \(L\) in equation~(\ref{mhd3})) are calculated explicitly, using a local fractional timestep technique to accurately treat those cells undergoing rapid temperature change. \New{Heating is principally due to absorption of stellar radiation by gas or dust: ionizing extreme ultraviolet (EUV) radiation in the ionized region, non-ionizing far ultraviolet (FUV) radiation in the neutral gas and x rays in the most shielded molecular gas. Cooling is principally due to collisionally excited line radiation of ions, atoms, or molecules. All of the various processes that contribute to \(H\) and \(L\), together with their implementation in the code, are discussed in detail in Appendix~\ref{sec:heating-cooling-laws}.} The individual components of the Phab-C\textsuperscript{2} code have been extensively tested against standard problems \citep{De-Colle:2005, 2004Ap&SS.293..173D, 2006MNRAS.371.1057I}. In addition, \citet{Arthur:2009} present further test cases for the combined algorithm.

\subsection{Initial and boundary conditions}
\label{sec:initial-conditions}

\begin{figure}
  \centering
  \includegraphics[width=\linewidth]{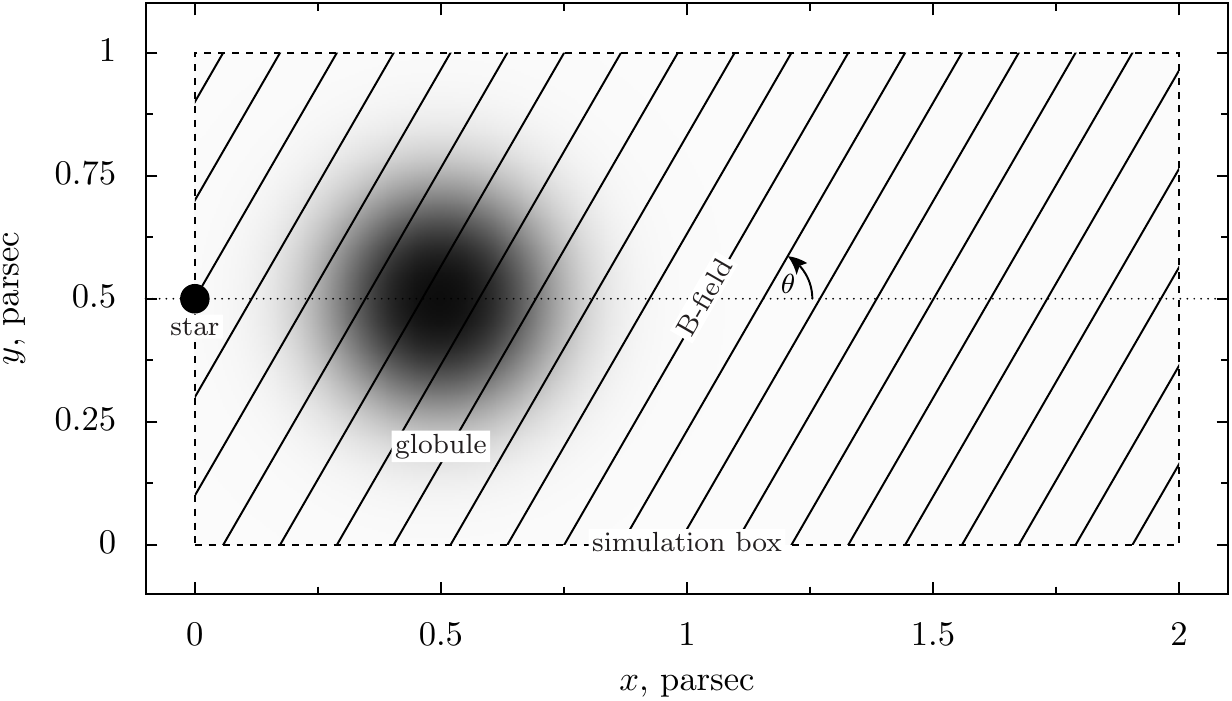}
  \caption{Initial conditions for the globule simulations. The
    grayscale shows the initial gas density in the midplane $z =
    0.5$~pc. }
  \label{fig:initial}
\end{figure}

The initial setup for our calculations is illustrated in Figure~\ref{fig:initial}.  Our simulations are calculated in a box of dimensions \( (x_\mathrm{max}, y_\mathrm{max}, z_\mathrm{max}) = (2, 1, 1)\)~pc.  The initial conditions consist of a spherical neutral globule with peak density \(n_0 = 10^4~\pcc\) and characteristic radius $r_\mathrm{g} = 0.2$~pc, located at a position \(\vec{r}_0 = (x_0, y_0, z_0) = (0.5, 0.5, 0.5)\)~pc. An ultraviolet radiation source with ionising photon luminosity \(Q_\mathrm{H} = 10^{49}~\mathrm{s}^{-1}\) and effective temperature \(T_\mathrm{eff} = 40,000\)~K is located at a position \(\vec{r}_* = (x_*, y_*, z_*) = (0.0, 0.5, 0.5)\)~pc, so that the initial distance of the globule centre from the source is \(D_0 = 0.5\)~pc. \New{The stellar FUV photon luminosity is \(Q_\mathrm{FUV} = 6.25\times 10^{48}~\mathrm{s}^{-1}\) and the stellar x-ray luminosity is assumed to be \(L_\mathrm{X,1} = 2.29 \times 10^{33}~\mathrm{erg\ s^{-1}}\) with temperature \(T_\mathrm{X,1} = 1.6 \times 10^7~\mathrm{K}\). In addition, we include a distributed x-ray source, corresponding to young low-mass stars in an accompanying star cluster, with a luminosity of \(L_\mathrm{X,2} = 3.47 \times 10^{33}~\mathrm{erg\ s^{-1}}\) and \(T_\mathrm{X,2} = 2.5\times 10^7~\mathrm{K}\), with the x-ray flux from this cluster calculated assuming a softening length of 0.3~pc.  These x-ray properties were chosen to be consistent with observations of the Trapezium cluster in the Orion Nebula \citep{2003ApJ...582..382F, 2005ApJS..160..379F, 2005ApJS..160..557S}. }
 
The globule has a smooth Gaussian density profile and is surrounded by a uniform ambient medium of density \(n_\mathrm{a} = 100~\pcc\), such that the initial density as a function of position \(\vec{r}\) in the box is given by 
\begin{equation}
  \label{eq:density}
  n(\vec{r}) = n_\mathrm{a} + (n_0 - n_\mathrm{a}) \exp( -\vert \vec{r} - \vec{r}_0 \vert^2 / r_\mathrm{g}^2) .  
\end{equation}
The initial ionisation fraction is set to the constant value of 0.01, while the initial temperature is set inversely proportional to the density, so as to give a constant pressure, with a temperature in the ambient medium of \(T_\mathrm{a} = 1500\)~K and a minimum temperature in the globule of \(T_0 = 15\)~K\@. The mean mass per hydrogen nucleon is set to \(\bar{m} = 1.3 m_\mathrm{p} = 2.174 \times 10^{-24}~\mathrm{g}\).

In this initial investigation of the problem, we vary only those parameters describing the magnetic field, keeping all other globule properties fixed. The globule mass for the adopted Gaussian density profile is \( M_\mathrm{g} = \pi^{3/2} r_{\mathrm{g}}^3 \bar{m} n_0 = 14.3~M_{\odot}\), whereas the mass of the diffuse ambient medium in our simulation box is \(M_{\mathrm{a}} = x_{\mathrm{max}} y_{\mathrm{max}} z_{\mathrm{max}} n_{\mathrm{a}} \bar{m} - n_{\mathrm{a}} M_{\mathrm{g}} / n_0 = 6.3~M_\odot\).

We assume that the magnetic field is initially uniform, with magnitude $B_0$, lying in the $xy$ plane at an angle $\theta_0$ to the $x$-axis (direction of ionising photons). We consider three strengths for the magnetic field: zero field (\(B_0 = 0\)), weak field (\(B_0 = 59~\mu\mathrm{G}\)), and strong field (\(B_0 = 186~\mu\mathrm{G}\)). The strong field is similar to the field strengths that have been observationally inferred in dark globules \citep[e.g.,][]{2003ApJ...592..233W}. The plasma \(\beta\) parameter (ratio of gas pressure to magnetic pressure) is initially constant throughout the simulation box, being \(\beta_0 = 0.1\) in the weak field case and \(\beta_0 = 0.01\) in the strong field case. Our most detailed investigations are of the \New{almost perpendicular field case (\(\theta_0 = 80^\circ\))}, but we also present simulations with \(\theta_0 = 0^\circ\) and \(\theta_0 = 45^\circ\). The three-dimensional simulations are carried out at resolutions of up to \New{\( 510 \times 255 \times 255\) }cells, and we also carry out one simulation in two-dimensional slab geometry at a resolution of \New{\(1002 \times 501\)} cells. The parameters for each run are summarised in Table~\ref{tab:models}. \New{We do not present simulations with (\(\theta_0 = 90^\circ\)) since the exact symmetry about the \(y = 0\) plane causes numerical difficulties at our lowest spatial resolution due to the formation of a thin dense sheet that is exactly aligned with the grid axes.}

\begin{table}
  \caption{
    Parameters for simulation runs
    \label{tab:models}
    }
  \begin{tabular}{lrrrrr}\hline
    Run & \(B_0\ (\mathrm{\mu G})\) & \(\beta_0\) &
    \(\theta_0 (^\circ)\) & Dim & \(N_x\) \\
    \hline
    S80H & 186 & 0.01 & 80 & 3 & 510 \\
    S80L & 186 & 0.01 & 80 & 3 & 254 \\
    S00L & 186 & 0.01 & 0 & 3 & 254 \\
    S45L & 186 & 0.01 & 45 & 3 & 254 \\
    W80L & 59 & 0.1 & 80 & 3 & 254 \\
    Z00L & 0 & \(\infty\) & 0 & 3 & 254 \\
    S80S & 186 & 0.01 & 80 & 2 & 1002 \\
    \hline
  \end{tabular}
\end{table}

\New{For all of these runs, standard outflow boundary conditions are adopted on all faces of the simulation cube. The effects of varying these boundary conditions are discussed in \S~\ref{sec:assessment}}

\section{Results}
\label{sec:results}

\begin{figure}
  \centering
  \includegraphics[width=\linewidth]{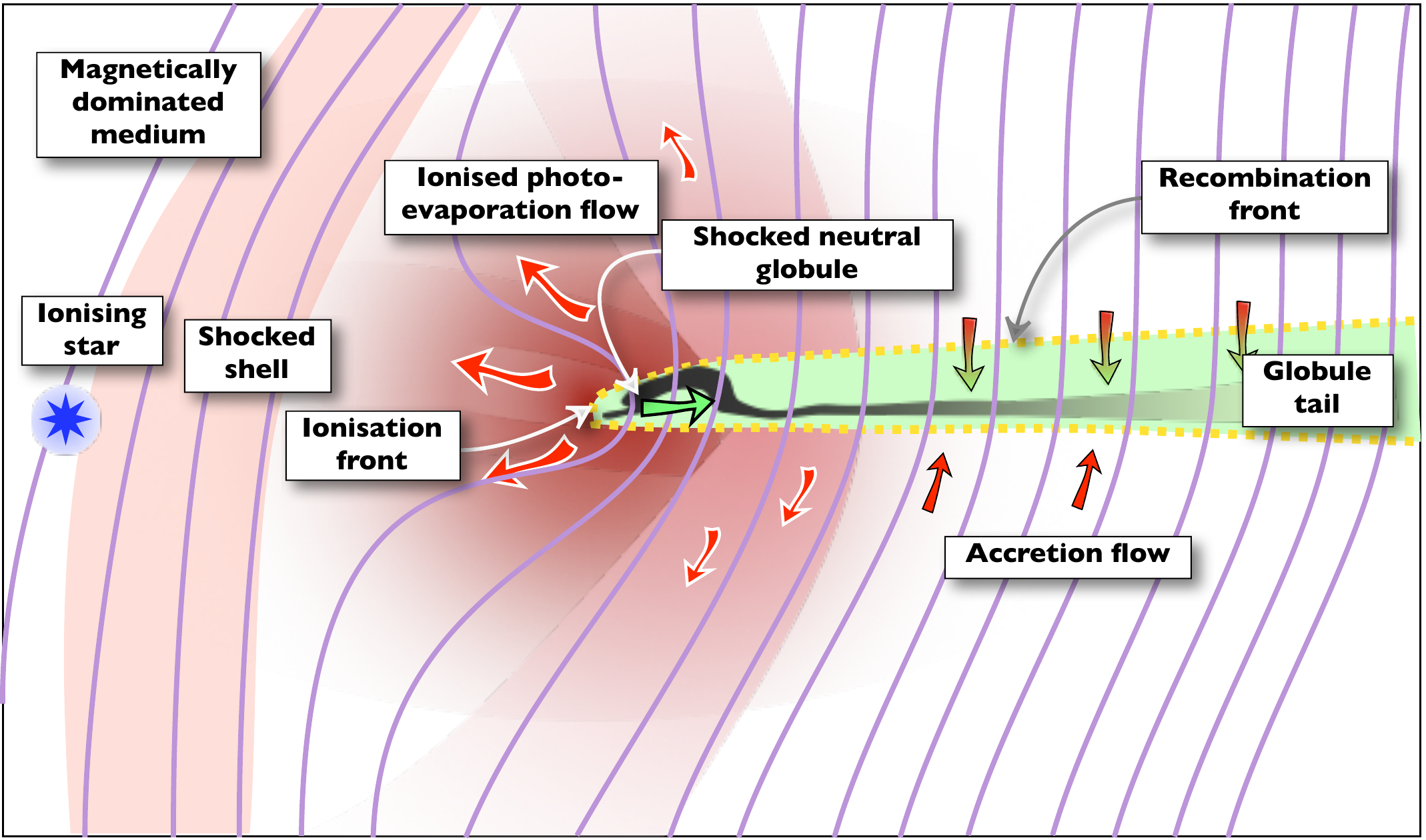}
  \caption{Idealised sketch of a photoevaporating magnetic globule.  Ionised gas is shown in red, neutral/molecular gas is shown in green, magnetic field lines are purple. The sketch shows the configuration for an initial magnetic field orientation of \New{\(\theta_0 \sim 80^\circ\)}, but qualitatively similar structures result from other field orientations. \label{fig:cartoon}}
\end{figure}

In the following description of the simulation results, particularly in figure captions, we will use a vocabulary of directions relative to the position of the globule, which should be interpreted as follows. ``Above'' and ``below'' refer to larger and smaller values of \(y\), respectively. ``Behind'' and ``in front'' refer to larger and smaller values of \(x\), respectively, so the ionising star is ``in front'' of the globule. ``To the side'' means away from the \(z = 0.5\)~pc symmetry plane.

Colour images of the optical appearance of the simulations are calculated as in \citet{2006ApJ...647..397M}, with each of the three colour channels representing the surface brightness in a different emission line: [\ion{N}{2}] 6584~\AA{} (red), H\(\alpha\) 6563~\AA{} (green), and [\ion{O}{3}] 5007~\AA{} (blue). Since our simulations do not explicitly follow the ionisation state of elements other than hydrogen, the ion fractions of N and O were approximated as fixed functions of the hydrogen ionisation fraction \citep{2005ApJ...621..328H}. The effects of dust absorption are included in calculating the images, asuming Orion Nebula dust properties \citep{1991ApJ...374..580B}. A sketch of a typical stage from our simulations is shown in Figure~\ref{fig:cartoon}, which explains some of the terms used in the description of our results. 

\subsection{Zero field: Z00L}
\label{sec:zero-field:-z00l}

\begin{figure}
  \centering
  \includegraphics[width=\linewidth]{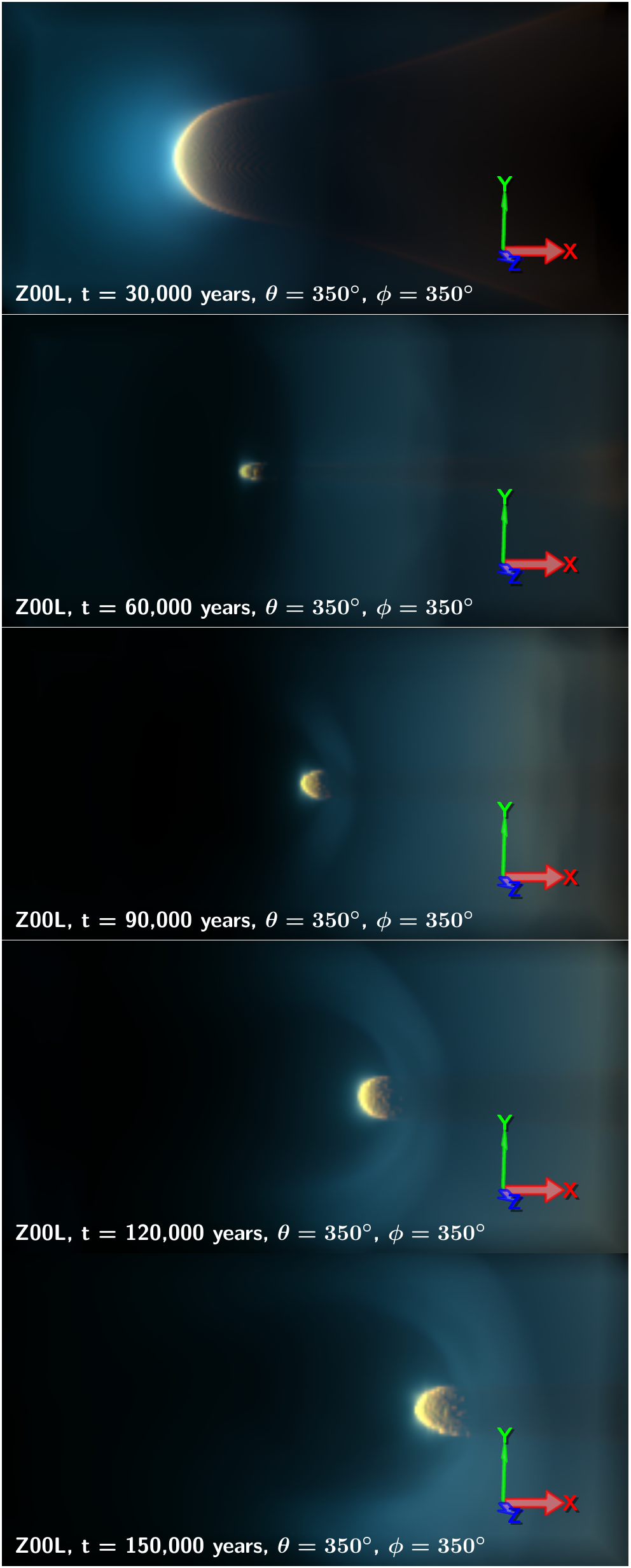}
 \caption[]{Sideways view \New{close to} the \(z\)-axis of the ionised emission from run~Z00L \New{for a sequence of} evolutionary times \New{from 30,000 to 150,000 years (top to bottom)}. The colours represent the surface brightness in three optical emission lines, as discussed in the text: [\ion{N}{2}] 6584~\AA{} (red), H\(\alpha\) 6563~\AA{} (green), and [\ion{O}{3}] 5007~\AA{} (blue). Thus, yellow-orange colours trace the ionisation front on the surface of the globule, while blue-green colours trace highly ionised gas. \label{fig:zero}}
\end{figure}

For reference, we first briefly discuss the globule evolution in the absence of a magnetic field (run Z00L), in which case, the globule evolution is cylindrically symmetric. \New{Various} snapshots of the evolution are shown in Figure~\ref{fig:zero}. Initially, the low density halo of the globule is ionised by a fast-moving R-type ionisation front, which transitions to D-type after about 1000 years, driving a convergent shock into the globule. The shock passes through the globule after about 50,000~years, after which time \New{of maximum compression (second panel in Fig.~\ref{fig:zero})} the globule is coherently accelerated by the back reaction of the transonic photoevaporated flow that leaves the ionisation front (rocket effect). At the same time, the ionised photoevaporated flow sweeps all the ambient material off the grid in the direction facing the star. \New{After the globule shock bounces off the symmetry axis, the neutral globule reexpands somewhat as it recedes from the star.}
% When the globule shock bounces off the symmetry axis, it triggers a series of acoustic oscillations in the shocked globule and tail, with a period of approximately 50,000~years. These oscillations excite travelling undulations in the ionisation front at the surface of the globule tail, which give rise to ``collars'' of enhanced density in the photoevaporation flow. 
All these features of the non-magnetic evolution have been well-studied by previous workers \citep{1989ApJ...346..735B, 1994A&A...289..559L, 1998A&A...331..335M,
  2001A&A...369..263P, 2001MNRAS.327..788W}. 

\subsection{Strong perpendicular field: S80H, S80L, S80S}
\label{sec:strong-perp-field}

\begin{figure*}
  \centering
  \includegraphics[width=\linewidth]{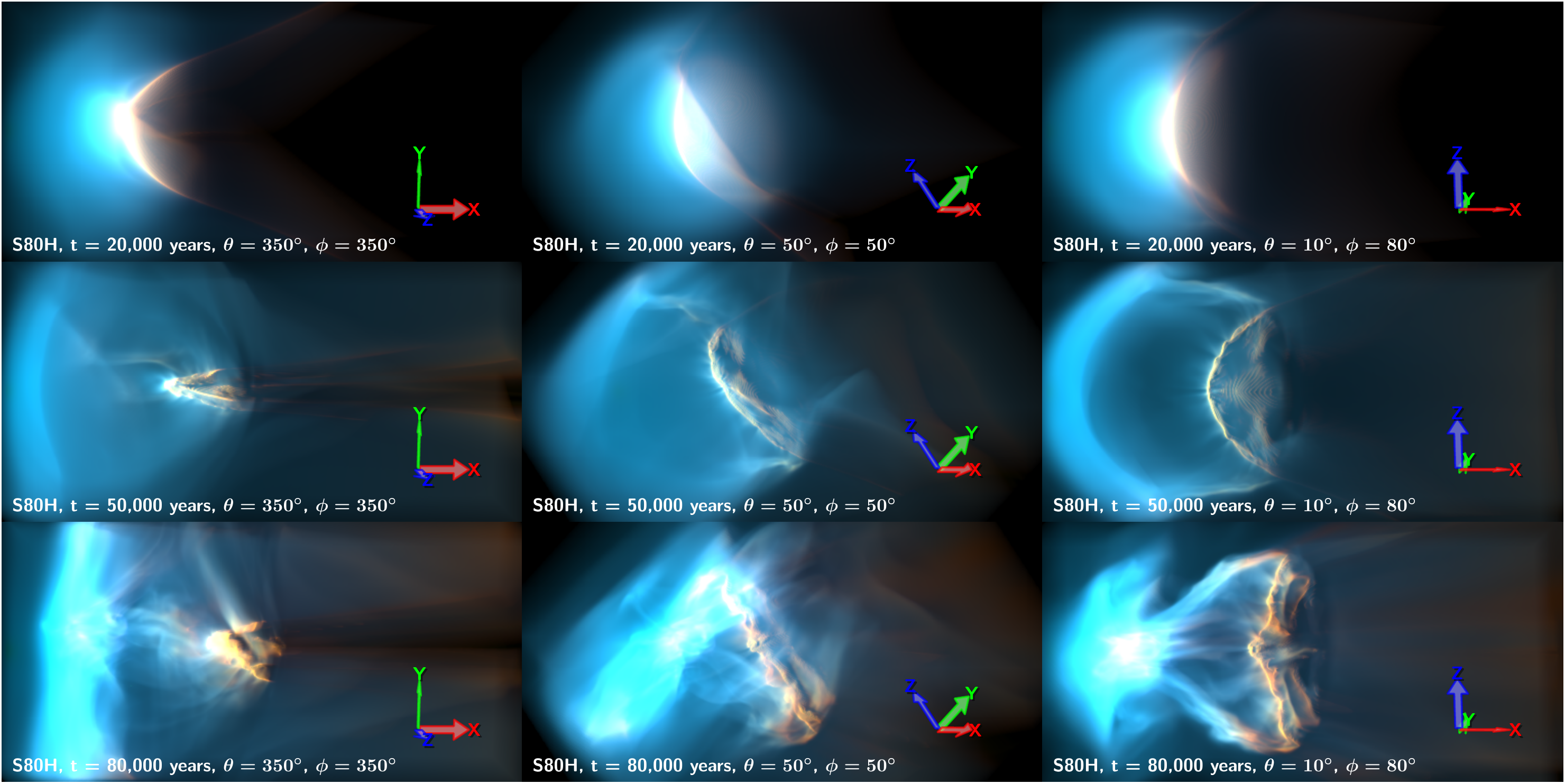}
  \caption[]{As Fig.~\ref{fig:zero} but showing multiple views of ionised emission from the \New{early} evolution of run~S80H (strong perpendicular field) \New{for 20,000 years (top), 50,000 years (middle), and 80,000 years (bottom).} \New{Left column shows the view from the side, slightly in front and slightly above. Central column shows the view from behind and below. Right column shows the view from below.} See \S~\ref{sec:numerical} for the definition of these in terms of the coordinate axes. }
  \label{fig:views}
\end{figure*}

In the ideal MHD approximation, matter is perfectly coupled to the magnetic field lines. Therefore, when the magnetic field in the neutral globule is very strong, magnetic pressure and tension strongly resist any movement of the gas in directions perpendicular to the original field direction, whereas no such magnetic support exists to oppose gas motions along the field lines. As a result, the implosion of the globule by the ionisation-driven shock front is highly anisotropic: the globule is swiftly flattened along the \(y\)-direction, whereas the radius of curvature in the \(xz\) plane remains large, giving a disk-like structure to the globule. This can be appreciated in Figure~\ref{fig:views}, which shows the structure of the ionised emission from run~S80H\@. At the earlier time shown (\New{20,000 years, top row}) the ionisation front already shows strong departures from cylindrical symmetry, such as the ``whiskers'' seen in the \(xz\) plane, while \New{after the point of maximum compression (50,000 years, middle row)}, the flattening is \New{considerable (aspect ratio of \(\simeq 5:1\))}.

A further difference from the non-magnetic case is seen in the behaviour of the ionised photoevaporation flow. Unlike in run Z00L, the photoevaporation flow cannot sweep all the ambient gas off the grid because the ambient magnetic pressure is too high. Instead, it forms a shocked shell, which is anchored by the large-scale magnetic field that threads the simulation box, as can be best seen as strong blue emission at the left-hand side of each panel in the \New{lower rows} of Figure~\ref{fig:views}.

\begin{figure*}
  \centering
 \includegraphics[trim=0 30 0 0, clip, scale=0.6] {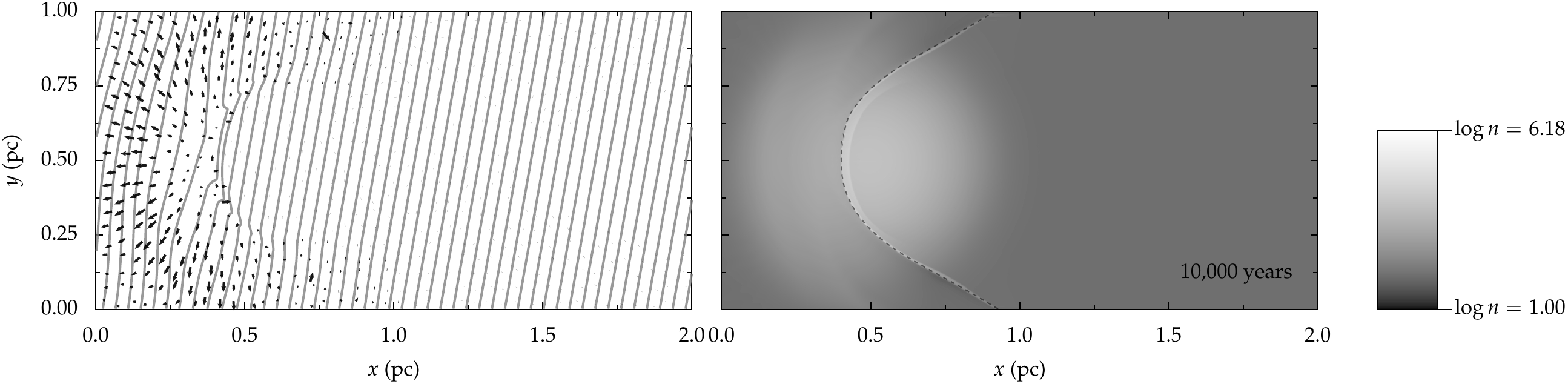}
  \includegraphics[trim=0 30 0 0, clip, scale=0.6] {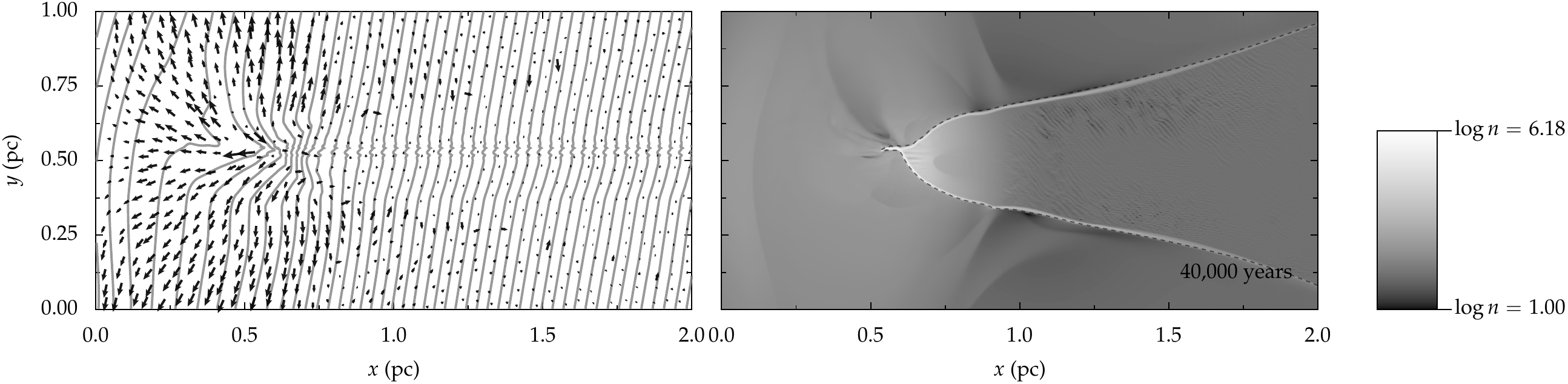}
  \includegraphics[trim=0 30 0 0, clip, scale=0.6] {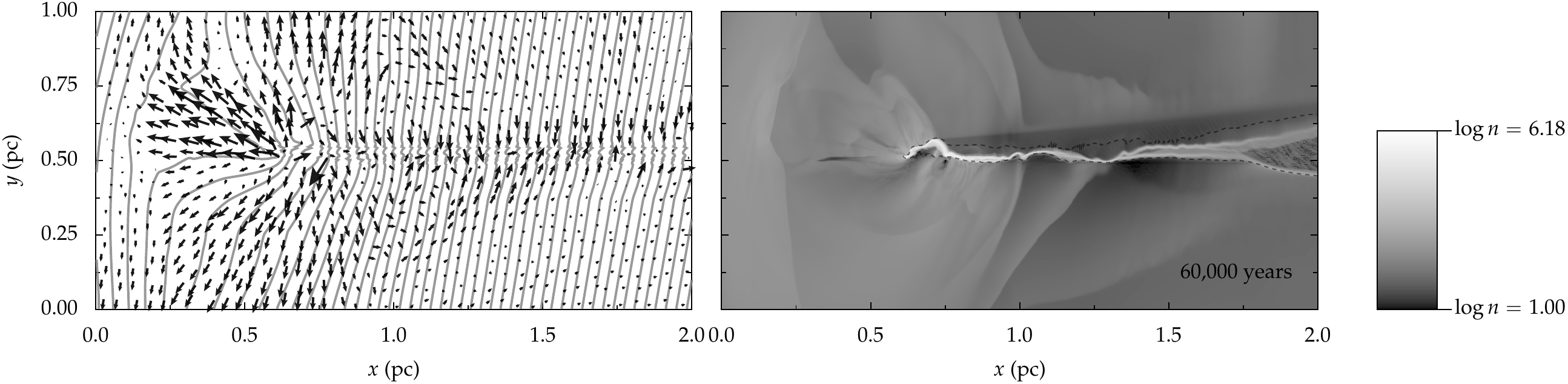}
  \includegraphics[trim=0 0 0 0, clip, scale=0.6] {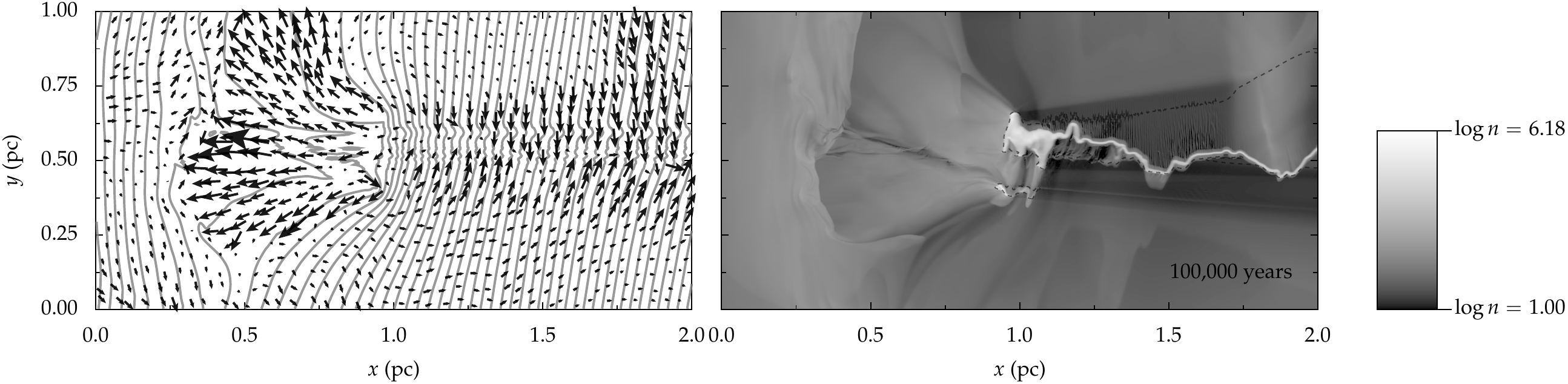}
\caption[]{
  Physical structure of run~S80S (strong perpendicular field, two-dimensional slab geometry) at evolutionary times of (top to bottom) \New{10,000, 40,000, 60,000 and 100,000~years}. \New{Left panels show magnetic field lines (grey lines) and gas velocity (black arrows). Right panels show logarithm of total hydrogen number density, \(n\) (grey scale, see scale bar) and ionisation fraction (dashed contour at a value of \(0.5\)). }
  }
  \label{fig:S80S}
\end{figure*}

In order to investigate the structure of the globule in greater detail, we have carried out a two-dimensional simulation in slab \(xy\) geometry (run S80S), which allows us to achieve a \(\sim 2\) times finer spatial resolution than in the highest resolution three-dimensional runs. The use of slab geometry is equivalent to imposing \( \partial / \partial z = 0\) for all quantities. This is a reasonably good approximation to the mid-plane (\(z = 0.5\)) of the three-dimensional runs with a strong perpendicular field, since the flattened form of the globule means that \(\partial / \partial z \ll \partial / \partial y\). The results are shown in Figure~\ref{fig:S80S} for three different times. Except where specifically noted, all the following description applies equally to the two-dimensional and three-dimensional run.

\subsubsection{Initial radiation-driven implosion}
\label{sec:early-evolution}

At the earliest time of 10,000 years (top \New{row} of Fig.~\ref{fig:S80S}), a fast-mode shock has started to run ahead of the ionisation front, compressing both the gas and magnetic field in the globule. This is followed by a slow-mode shock, which at this earliest time can be seen just starting to detach from the ionisation front at the sides of the globule. The ionised portion of the globule is starting to flow back towards the star due to the pressure imbalance (the equilibrium temperature in the ionised gas is \New{almost} independent of density), but the photoevaporation flow is not yet fully developed.

At the intermediate time of \New{40,000} years (\New{second row} of Fig.~\ref{fig:S80S}), both the fast-mode and slow-mode shock are clearly visible, which compress the gas by a factors of \New{\(\sim 2\) and \(\sim 10\), respectively}, producing a neutral shell with density \New{\(\sim 20\)} times greater than the initial peak globule density. In addition, the slow-mode shock is of the ``switch-off'' type, in which the post-shock magnetic field has no component in the plane of the shock. The dense neutral shell is therefore effectively demagnetised and gas-pressure dominated (\(\beta \sim 5\)). Since the neutral shell is thin, the ionisation front and the slow-mode shock are approximately parallel, meaning that the B-field is perpendicular to the ionisation front over much of the surface of the globule. Such fronts, in which the jump conditions reduce to the pure hydrodynamic case, have been dubbed ``extra strong'' \citep{2000MNRAS.314..315W, 2007Ap&SS.307..179W} since the flow passes through both the Alfv\'en and sound speeds.

The slow magnetosonic speed is zero in the direction exactly perpendicular to the magnetic field lines, so the slow-mode shock cannot propagate in the symmetry plane (\(y = 0.5~\mathrm{pc}\)). Instead, a dense ridge is formed from the pincer-like convergence of the neutral shell from above and below. The ridge accretes both material and horizontally oriented field through fast-mode shocks at its upper and lower boundaries. \New{The triply-shocked gas in the ridge has densities as high as \(2 \times 10^6~\pcc\), 200 times denser than the initial globule.} The field lines from above and below the midplane have opposite directions, so a current sheet forms in the mid-plane, in which magnetic reconnection can occur.

% \TODO{More about reconnection here, or possibly in an appendix. We could do with a diagram to make things clearer. Sweet-Parker versus Petschek modes for reconnection. Monitor flux loss in the simulations. Analytic estimate of rate at which flux is advected into the current sheet. Discussion of numerical magnetic diffusivity. Magnetic flux destruction rate should be independent of grid resolution. Reconnection slab jet squirts into the ionised region. Is it already super-Alfv\'enic on the neutral side? This seems necessary to have supersonic flow off the i-front with a non-zero transverse field.}

\subsubsection{Accelerated globule phase}
\label{sec:accel-globule}

By \New{a time of 60,000 years (third row of Fig.~\ref{fig:S80S})}, the initial implosion shocks have passed through the \New{entire} globule, \New{forming a thin, dense sheet of material in the midplane.} The neutral globule has a much more complicated internal structure than in the pure-hydrodynamic case due to the mutual interactions of the multiple shocks that have reflected from the midplane. However, the head of the globule has a configuration that is qualitatively similar to that seen at the earlier time: the magnetic field lines are approximately perpendicular to the ionisation front everywhere except for the nose that faces the star. The neutral dense ridge that was forming at \New{40,000} years has by now been completely ionised, so that the nose now consists of material from the neutral shell that formed on the flanks of the original globule. The magnetic field lines in these parts of the shell were not bent far from their initial \(y\)-orientation,\footnote{The effects of the oblique fast-mode and slow-mode shocks are in opposite directions and approximately cancel.} so the mid-plane current sheet has disappeared. \New{Thin-shell instabilities start to grow in the shocked sheet. Although these instabilities occur in all our simulations, their onset is slightly earlier and their growth is more vigorous for the higher resolution runs.}
% \TODO{Mention small differences between 2d and 3d simulations?}

The ionised photoevaporation flow from the head of the globule at this time consists of two zones with distinct properties. The flow from the flat nose of the globule is very weakly magnetised (\(\beta = 10\)-\(100\)), whereas the surrounding sheath flow from the globule shoulders is closer to equipartition (\(\beta = 1\)--\(3\)). For comparison, after ionisation by an R-type front, the quiescent ambient medium would have \(\beta \simeq 0.2\). This structure does not remain constant in time, but evolves as the ionisation front progresses through different regions of the complex magnetic geometry in the globule head, with each zone of the photoevaporation flow slowly peeling back towards the sides. In both zones the flow is mildly supersonic and accelerating, reaching maximum velocities \(\simeq 30~\kms\). Fine-scale structure in the magnetic field at the ionisation front produces ``streamers'' in the photoevaporation flow, which typically have lower-than-average field strength, together with higher-than-average density and velocity. The whole photoevaporation flow is criss-crossed with weak oblique shocks.

The shadow tail behind the globule is accreting gas along the field lines from the ambient medium to either side, due to the higher thermal pressure in the ionised gas. This accretion flow, with velocity \(\simeq 10~\kms\), recombines in a broad front as it enters the region shielded from direct ionising radiation, and then shocks against the \New{thin} dense (\New{\(n \simeq 10^4\)--\(10^5~\pcc\)}) core of the tail. It should be noted that our treatment of the diffuse ionising field is very approximate and that this will predominantly effect the structure of the tail region \citep{1998ApJ...502..695C}. However, globule simulations that include an exact treatment of the diffuse field \citep{Raga:2009} show that \New{the differences are not large in most cases.}

\New{At later times, the instabilities in the shocked globule continue to grow. In the globule head, these cause fragments of dense neutral gas to separate from the main globule, forming isolated smaller globules, as can be seen at at time of 100,000~years in Figure~\ref{fig:S80S}. In the globule tail, the instabilities cause portions of the shocked sheet to exit the zone behind the head that is shadowed from direct ionising radiation. When this happens, the tail portion begins to be photoevaporated, but this is a transient phenomenon since the rocket effect then drives the neutral sheet back into the shadow zone. The process repeats when the sheet leaves the other side of the shadow, producing quasi-periodic oscillations on a timescale of tens of thousands of years.   }

\subsubsection{Late-time recombination of the shocked shell}
\label{sec:late-evolution:-s90l}
\begin{figure*}
  \centering
  \includegraphics[width=\linewidth]{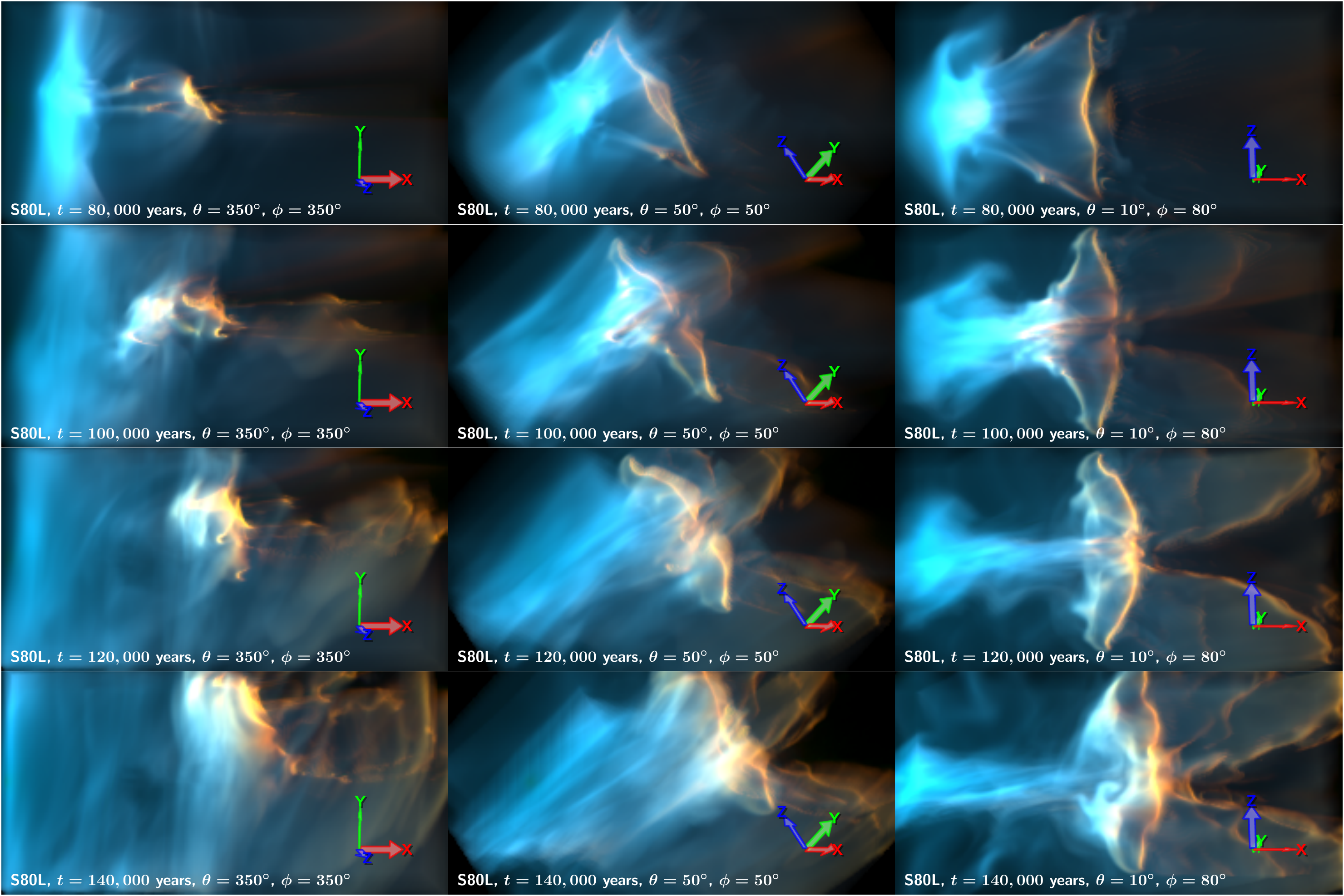}
  \caption{\New{As Fig.~\ref{fig:views} but showing multiple views (left, centre and right columns) of the late-time evolution of a run~S80L (strong, nearly perpendicular field) at times from 65,000 (top row) to 140,000 years (bottom row). The first row is for a time shortly after the last time shown in Fig.~\ref{fig:views} and illustrates a weaker growth of instabilities in the shocked globule in this lower resolution run. The second row shows the moment in which material in the dense shocked ribbon and cavity walls have recombined to shadow sections of the original globule (see text), whereas, by the time of the third row, this material has been reionised. At the time of the fourth row, the globule is about to leave the top of the grid after being deflected by the magnetic field.}}
  \label{fig:ribbon}
\end{figure*}

At times later than 60,000~years, we begin to see significant deviations between the behaviour of the two-dimensional and three-dimensional runs.\footnote{We have only been able to \New{fully} investigate this late-time evolution in the lower resolution run S80L.} This is due to the evolution of the shocked shell that forms by the interaction of the ram-pressure dominated photoevaporation flow with the magnetically dominated ambient medium.  In the three-dimensional run, the cavity that is swept out by the photoevaporation flow has a finite size in the \(z\) direction of about 0.7~pc, roughly equal to the globule radius of curvature in the \(xz\) plane, whereas in the two-dimensional run, the cavity is implicitly unlimited in its \(z\) extent. As the globule is pushed away from the star by the rocket effect, the flow cavity moves with it, vacating the region with \(x < 0.5~\mathrm{pc}\). In the three-dimensional run, the ambient medium that was pushed aside by the photoevaporation flow then closes in behind the cavity like curtains, driven by the magnetic pressure and tension in the cavity walls. At a time of \(\simeq 80,000\)~years, the two sides of the shocked shell collide with one another in the symmetry plane \(z = 0.5\)~pc at a relative velocity of \(\simeq 40~\kms\), forming a dense ribbon (\(n \simeq 1500~\pcc\)) of ionised gas, elongated along the \(y\)-axis. \New{This can be seen in the bottom row of Figure~\ref{fig:views} for the high-resolution run S80H and in the top panel of Figure~\ref{fig:ribbon} for the low-resolution run S80L\@. }

A few thousand years later, the density in the shocked ribbon is sufficiently high to trap the ionisation front, causing \New{parts of} the ribbon to recombine, as do sections of the cavity wall. After this time, the central part of the original globule is in the ionisation shadow of the ribbon, so this too recombines, the photoevaporation flow ceases, and the now-overpressured neutral globule begins to expand and dissipate. \New{This can be seen in the second row of Figure~\ref{fig:ribbon}. However, by a time of 120,000 years (third row of Fig.~\ref{fig:ribbon}), all of this recombined material has been reionised and photo-evaporated, so that the original globule is once again illuminated by the ionising radiation. At the same time, as with the 2D models discussed above, the globule's trajectory begins to veer increasingly away from the \(x\)-axis as its motion becomes closer to parallel to the magnetic field lines. The complex flow structure in this late stage is seen in the bottom row of Figure~\ref{fig:ribbon}.}

% At the same time, the neutral ribbon begins to fragment into a string of knots of density \(\simeq 4000~\pcc\) and mass \(0.1\)--\(1~M_\odot\), each of which will now be photoevaporated in a manner similar to the original globule.\footnote{The neutral parts of these knots have \(\beta \sim
%   1\) so magnetic effects are not very important in their subsequent
%   evolution.} The complex flow structure at a time of 90,000~years is
% shown in Figure~\ref{fig:ribbon}.

\New{
\subsubsection{Thermal behavior of the globule}
\label{sec:therm-behav-glob}
\begin{figure*}
  \centering
  \setkeys{Gin}{trim=25 30 0 0, clip=on}
  \includegraphics[trim=0 30 0 0, scale=0.9]{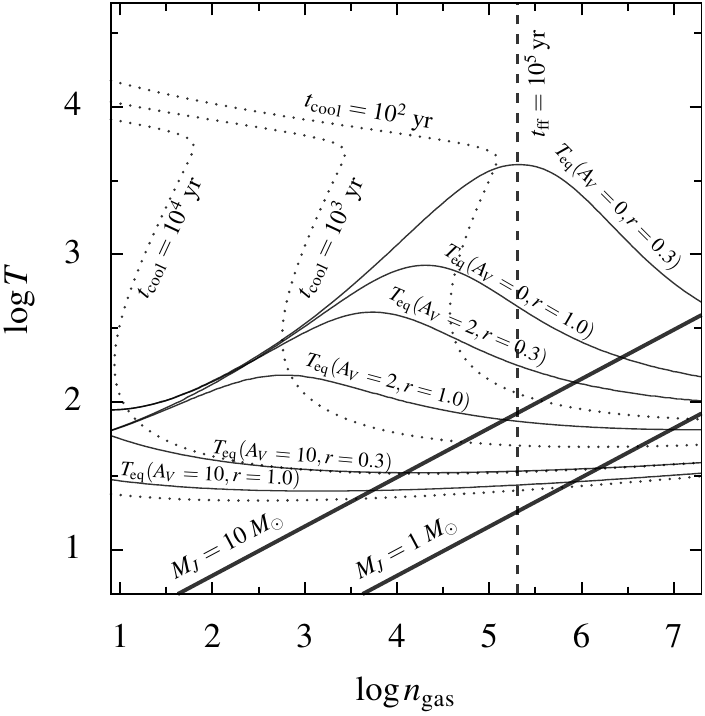}%
  \includegraphics[clip, scale=0.9]{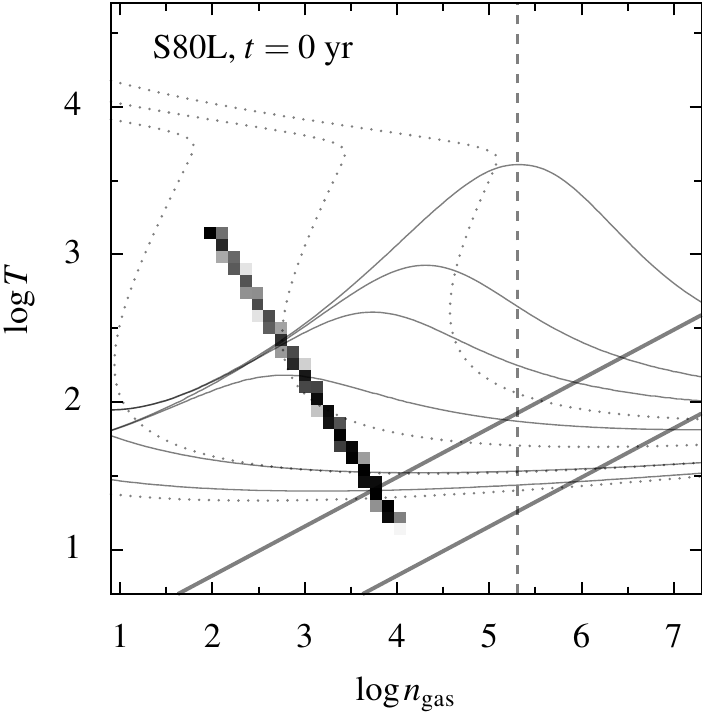}%
  \includegraphics[clip, scale=0.9]{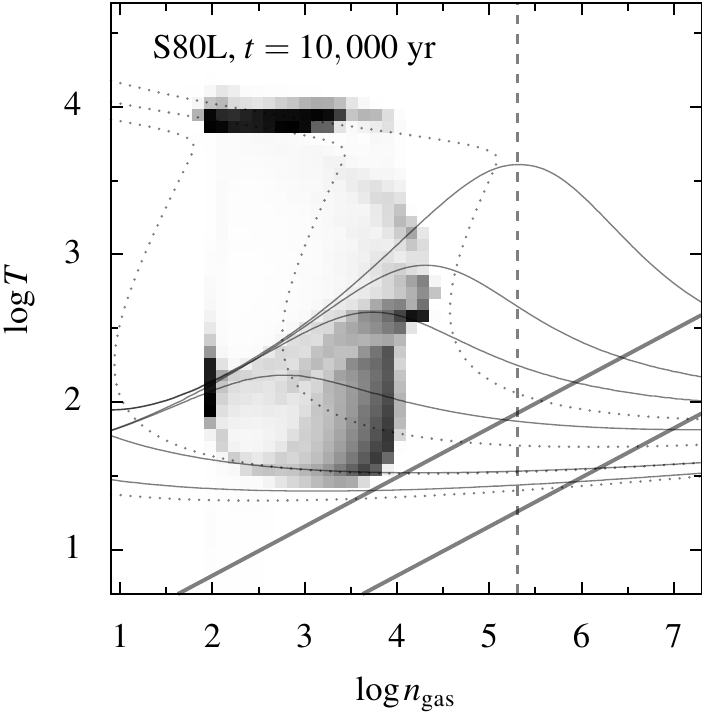}\\
  \includegraphics[trim=0 0 0 0, clip, scale=0.9]{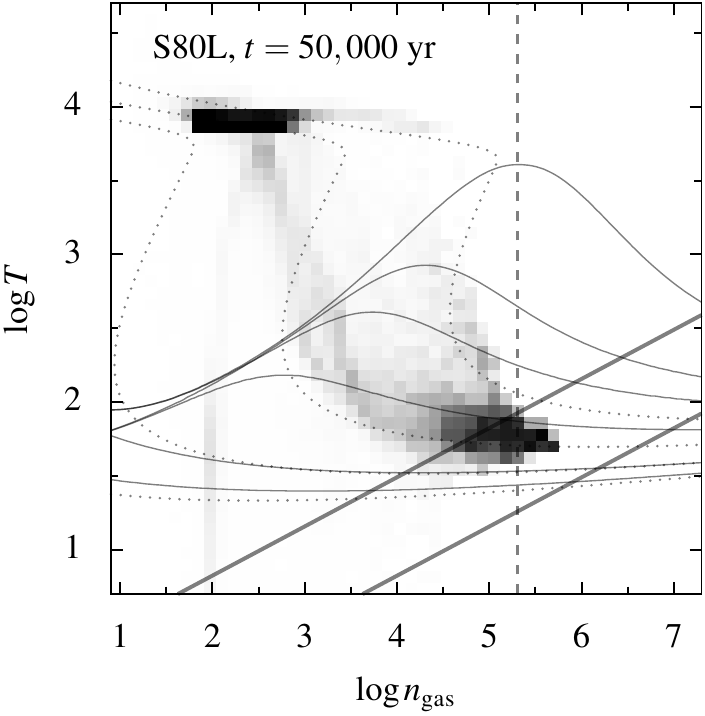}%
  \includegraphics[trim=25 0 0 0, clip, scale=0.9]{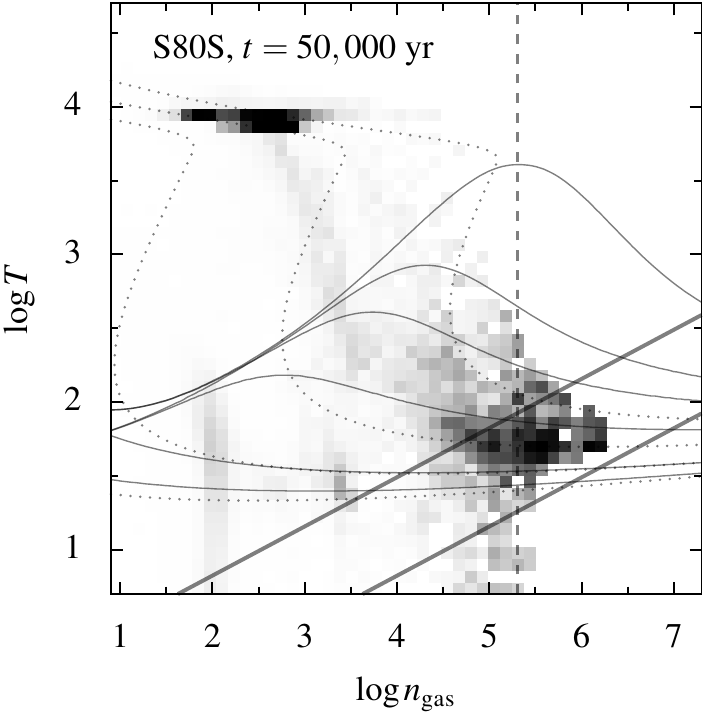}%
  \includegraphics[trim=25 0 0 0, clip, scale=0.9]{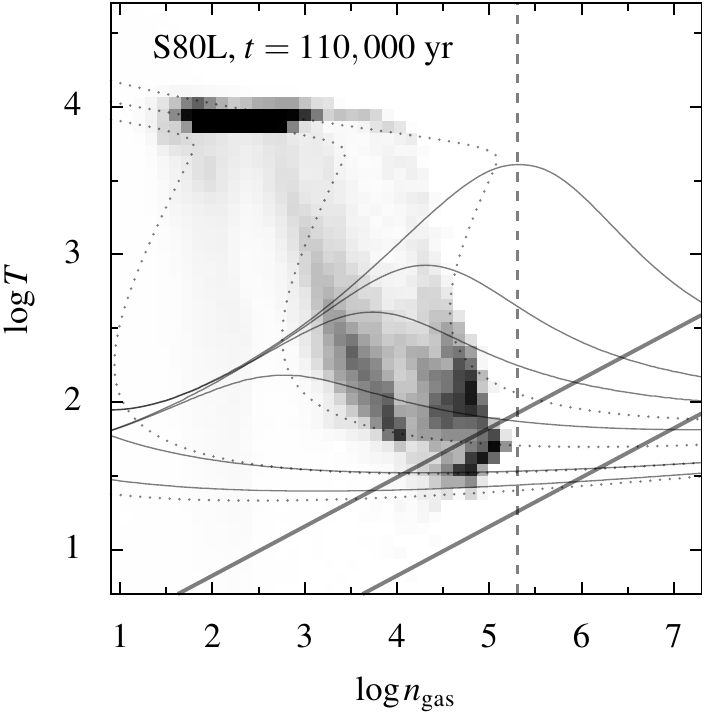}%
%   \makebox[0pt][r]{\raisebox{0.167\linewidth}{\framebox{\textbf{To be replaced by a later one}}\quad}}%
  \caption{\New{Evolution of thermodynamic quantities. The grayscale images show the fraction of the total gas mass in the simulation that lies within a given logarithmic interval of density (\(x\) axis)  and temperature (\(y\) axis). Results are shown for a sequence of evolutionary times of model S80L\@. Thin solid lines show equilibrium temperatures for neutral/molecular gas, at different distances, \(r\) in parsecs, and extinguishing columns, \(A_V\) in magnitudes, from the ionising star---top to bottom: \((A_V, r) = (0, 0.3)\), \((0, 1.0)\), \((2, 0.3)\), \((2, 1.0)\), \((10, 0.3)\), and \((10, 1.0)\).  Dotted lines show contours of radiative cooling time---top-right to bottom-left: \(t_{\mathrm{cool}} = 100\), \(1000\),  \(10,000\), and \(100,000\)~years. Heavy diagonal solid lines indicate Jeans masses of \(10~M_\odot\) (upper line) and \(1~M_\odot\) (lower line)---Jeans instability requires a clump of the requisite mass to be below the line. The region to the right of the vertical dashed line corresponds to a gravitational free-fall timescale \(< 100,000\)~years.}}
  \label{fig:n-T}
\end{figure*}

Figure~\ref{fig:n-T} shows the evolution with time of the gas density and temperature for model S80L\@. The initial conditions are shown in the top-centre panel, appearing as a diagonal line corresponding to \(n T = \mathrm{constant}\), since the initial configuration is at constant pressure. This initial configuration is not in thermal equilibrium, mostly lying above the region delimited by the PDR equilibrium curves (thin solid lines in the figure), and with a cooling time (dotted lines in the figure) of a few thousand years.

By a time of 10,000 years (top-right panel), the initial configuration has been modified in various ways. A portion of the gas has been ionised, giving the horizontal band at roughly constant temperature (\(\log T \simeq 3.9\)). Another portion of the gas lies in the thin cooling zone behind the convergent fast-mode shock that is driven into the globule, giving the curved band of material between \((\log n, \log T) = (3.0, 2.0)\) and \((4.1, 2.6)\). The small amount of material around \((\log n, \log T) = (4.0, 3.1)\) has been shocked by the slow-mode shock that at this time has just detached from the ionisation front on the flanks of the globule (see Fig.~\ref{fig:S80S}). The remainder of the globule gas (\(\log n = 2.5\) to \(4.0\)) has reached its equilibrium PDR temperature of \(\log T = 1.5\) to \(2.3\), depending principally on the extinguishing column, which at this early stage of the implosion is still moderate (\(A_V \simeq 3\) to the center of the globule). 

By a time of 50,000~years (bottom-left panel), the core of the globule reaches its maximum compression, having been reshocked to densities \(\sim 10^6~\mathrm{cm}^{-3}\). Due to the 4 times lower linear resolution, this is about a factor of 3 lower than the maximum density seen in the slab-symmetric two-dimensional model S80S discussed in \S~\ref{sec:early-evolution}. The density-temperature distribution of model S80S at \(t = 50,000\)~years is also shown in Figure~\ref{fig:n-T} (bottom-centre panel). It is very similar to S80L, except for extending to slightly higher densities in the imploded globule. The greater noise apparent in the S80S graph is due to the fact that, despite the higher resolution, it has considerably fewer grid cells than S80L. 

Our simulations do not include self-gravity, but an estimate of whether gravitational effects may be important can be obtained by considering the Jeans instability criterion and the free-fall time. The heavy solid lines in Figure~\ref{fig:n-T} show the threshold for Jeans instability for globules of \(10~M_\odot\) and \(1~M_\odot\): \(M_J = 2 M_\odot (c/0.2~\mathrm{km\ s^{-1}})^3 (n/10^3~\mathrm{cm^{-3}})^{-1/2}\). This can only be taken as indicative, since it does not include the effects of magnetic fields nor turbulent motions, which would help suppress the instability. The total neutral mass of the globule at this point is about \(6~M_\odot\), but only a fraction of this mass has the highest densities. Furthermore, the dense gas is not in a compact configuration, but rather is in the form of a ridge, which would further reduce its effective mean density for the purposes of triggering the instability, so it seems likely that the globule would resist global gravitational collapse. An additional argument is that the free-fall time (\(\propto n^{-1/2}\)) is of order \(100,000\)~years, which is long compared with the dynamical evolution time. It may be that a fraction of the globule mass is compressed to such high densities that it undergoes prompt gravitational collapse, but addressing this question would require a much higher spatial resolution than we can achieve in our fixed-grid simulations. 

By  a time of 110,000~years (bottom-right panel), the globule has rebounded from its maximum compression and entered its equilibrium acceleration phase (\S~\ref{sec:accel-globule}). At this point, the density-temperature distribution consists of a horizontal band of ionised gas at \(\log T \simeq 3.9\), from which descend 3 strands of neutral material towards lower temperatures and higher densities. The rightmost strand corresponds to the bulk of the original globule material, the central strand corresponds to the shocked tail, whereas the leftmost strand, which is very faint since it contains little mass, corresponds to the partially ionised transonic accretion flow onto the tail. 

It is obvious from Figure~\ref{fig:n-T} that the widely used assumption of an isothermal equation of state for the neutral/molecular gas is a very poor approximation for much of the evolution of the globule. The combined effect of heating by shocks and stellar x~rays keeps the bulk of the dense gas (\(> 10^5~\pcc\)) at temperatures in the range 50 to 100~K, whereas less dense neutral gas (\(\sim 10^4~\pcc\)) is frequently heated to 100 to 500~K by FUV radiation. This will have important consequences for questions of gravitational instability and triggered star formation. For instance, \citet{2007MNRAS.377..383E} assumed a constant \(T = 10\)~K for the neutral gas in simulations of the photoevaporation of self-gravitating clumps, finding that the Jeans instability was triggered in some cases. If they had used \(T = 50\)~K, then the Jeans mass would have been \(5^{3/2} = 11.2\) times higher and their clumps would have remained gravitationally stable. 
} % End New

\Newest{One shortcoming of our simulations is that we use one-dimensional models to calibrate the heating and cooling functions (Appendix~\ref{sec:heating-cooling-laws}), whereas in reality multidimensional effects may be important. In particular, during its initial compression phase the neutral globule is compressed to a thin sheet with a high optical depth to stellar photons, which propagate parallel to the sheet, but a much lower optical depth in the perpendicular direction. This may enhance the cooling efficiency in the dense gas by allowing optical and near-infrared radiation to escape more easily. However, in practice this effect is likely to be small since the cooling for \(T < 1000\)~K is dominated by far-infrared and millimeter lines, for which the sheet is optically thin in all directions.}

\subsection{Effects of varying the initial field angle}
\label{sec:variation-with-field}

\begin{figure}
  \centering
  \includegraphics[width=\linewidth]{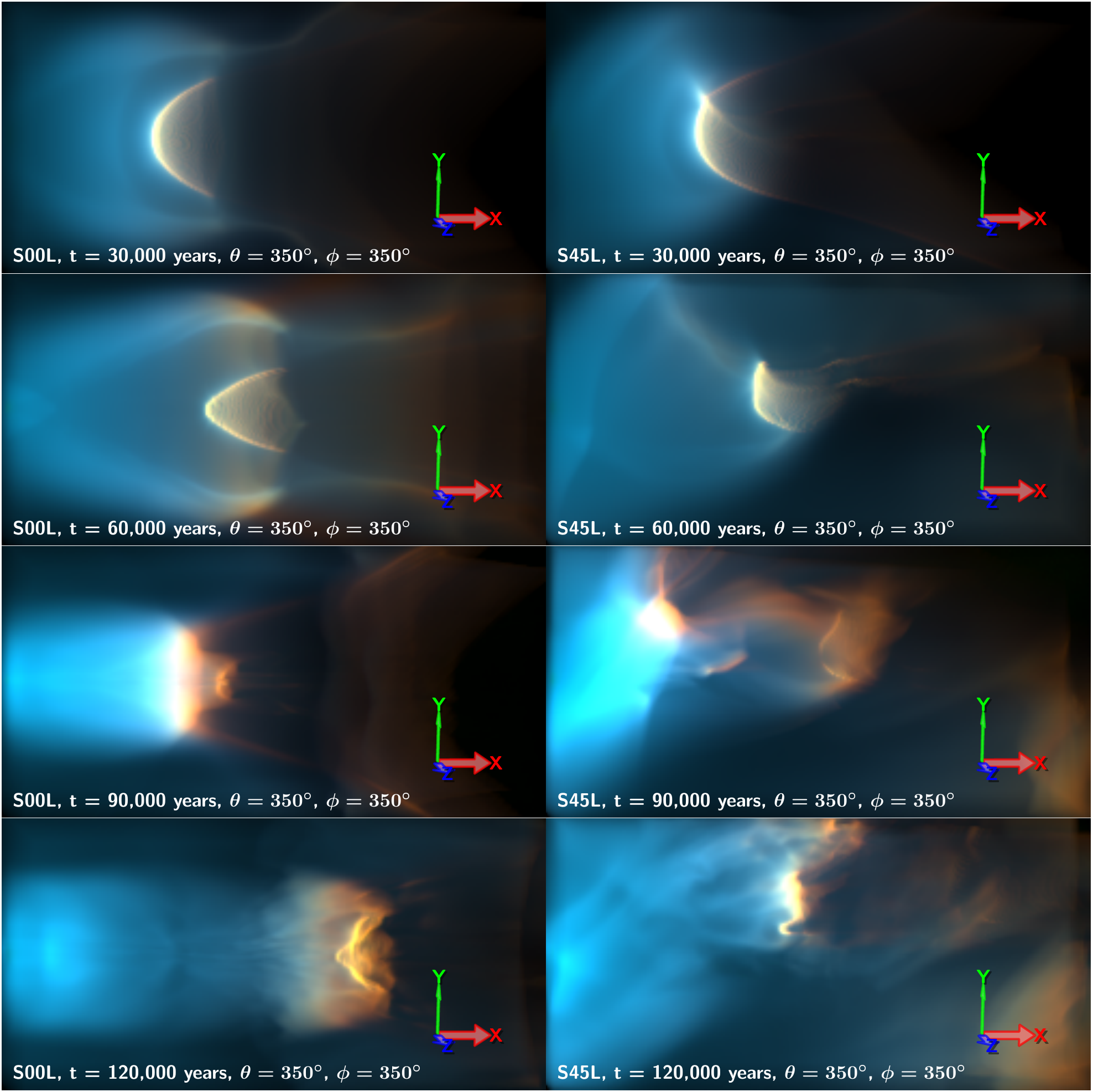}
  \caption{As Fig.~\ref{fig:zero} but showing views of ionised emission from runs~S00L (strong parallel field, left column) and S45L (strong slanted field, right column), for a sequence of evolutionary times. View directions are the same as in the leftmost column of Fig.~\ref{fig:views}. }
  \label{fig:views-00-45}
\end{figure}

In order to investigate the effect of the initial magnetic field geometry on the globule evolution, we have carried out runs with \(\theta_0 = 45^\circ\) and \(0^\circ\). These are both at the lower resolution of \(255 \times 127 \times 127\).

\subsubsection{Strong parallel field: S00L}
\label{sec:strong-parall-field}

When \(\theta_0 = 0^\circ\), the \(x\)-axis, \((y, z) = (0.5, 0.5)~\mathrm{pc}\), becomes a symmetry axis of the problem, as in the non-magnetic case, so that the globule remains cylindrically symmetric throughout its evolution. One important difference with respect to the non-magnetic case is that the support provided by the magnetic field opposes the lateral compression of the neutral globule, producing a broader, snubber globule head. The internal structure of the compressed globule is much simpler than in the perpendicular field case: consisting of a thin shell behind the ionisation front, with \New{peak density \(n \simeq 10^6~\pcc\)}, plus a lower density core with \(n \simeq 5000~\pcc\). The shell is \New{moderately magnetically} dominated \New{(\(\beta \simeq 0.3\))}, while the core \New{and the tail (which maintains its ambient density of  \(100~\pcc\)) are overwhelmingly magnetically dominated (\(\beta \simeq 0.001\))}. The field lines within the globule never get bent through large angles and there is no equivalent of the current sheet that forms in the \(\theta_0 = 80^\circ\) runs. There is no accretion flow into the tail because the thermal overpressure of the ionised gas at the sides of the globule is not sufficient to laterally compress the tail's longitudinal magnetic field. The ionised emission from the S00L run \New{for various evolutionary times} is shown in the left column of Figure~\ref{fig:views-00-45}.

The magnetic field in the shocked neutral shell is approximately perpendicular to the globule surface over the entire globule head, so that ionisation front is again of the ``extra-strong'' type. The ambient magnetic field channels the photoevaporation flow towards the symmetry axis by means of a focusing shock. At around \New{60,000~years} the sides of the shocked shell become optically thick to ionising radiation \New{(shown in the second row of Fig.~\ref{fig:views-00-45})}, and a few thousand years later so does a dense ionised knot that has formed on the symmetry axis. This causes the recombination of the photoevaporation flow \New{(third row of Fig.~\ref{fig:views-00-45})} in a similar way to the \(\theta_0 = 80^\circ\) runs (\S~\ref{sec:late-evolution:-s90l}), \New{although the recombination period lasts much longer in this case (bottom two rows of Fig.~\ref{fig:views-00-45}). }

\subsubsection{Strong slanted field: S45L}
\label{sec:slanted-field:-s45l}

The evolution of the \(\theta_0 = 45^\circ\) run shares elements of both the \(\theta_0 = 80^\circ\) (\S~\ref{sec:strong-perp-field}) and \(\theta_0 = 0^\circ\) runs (\S~\ref{sec:strong-parall-field}), but also has unique features of its own. The ionised emission from the S45L run is shown in the right column of Figure~\ref{fig:views-00-45} for a \New{sequence of evolutionary times}. The problem no longer has an \(xz\) symmetry plane and the globule acceleration is not strictly radial away from the ionising star. The magnetic field is not \New{initially} strong enough to force the globule to move like a bead on a wire, but magnetic tension forces are sufficient to deflect the globule motion by \(\simeq 15^\circ\) from the \(x\)-axis.  A feature analogous to the dense ridge that formed in the \(\theta_0 = 80^\circ\) runs at the symmetry plane now forms at the top corner of the globule head, where the initial magnetic field is tangential to the globule surface. However, unlike in the \(\theta_0 = 80^\circ\) case, the ridge does not have a large radius of curvature in the \(xz\) plane and the globule does not become significantly flattened. Instead, the unbalanced forces due to the one-sided ionising illumination of the ridge cause it to ``curl up'' away from the ionising star, so that the magnetic geometry of the head becomes similar to that in the \(\theta_0 = 0^\circ\) run, with the photoevaporation flow streaming off the ionisation front along approximately radially oriented field lines. As with the other strong field runs, the shocked ionised shell eventually traps the ionisation front, leading to recombination of the photoevaporation flow and a complex subsequent evolution.

\subsection{Weak perpendicular field: W80L}
\label{sec:weak-perp-field}

\begin{figure*}
  \centering
  \includegraphics[width=\linewidth]{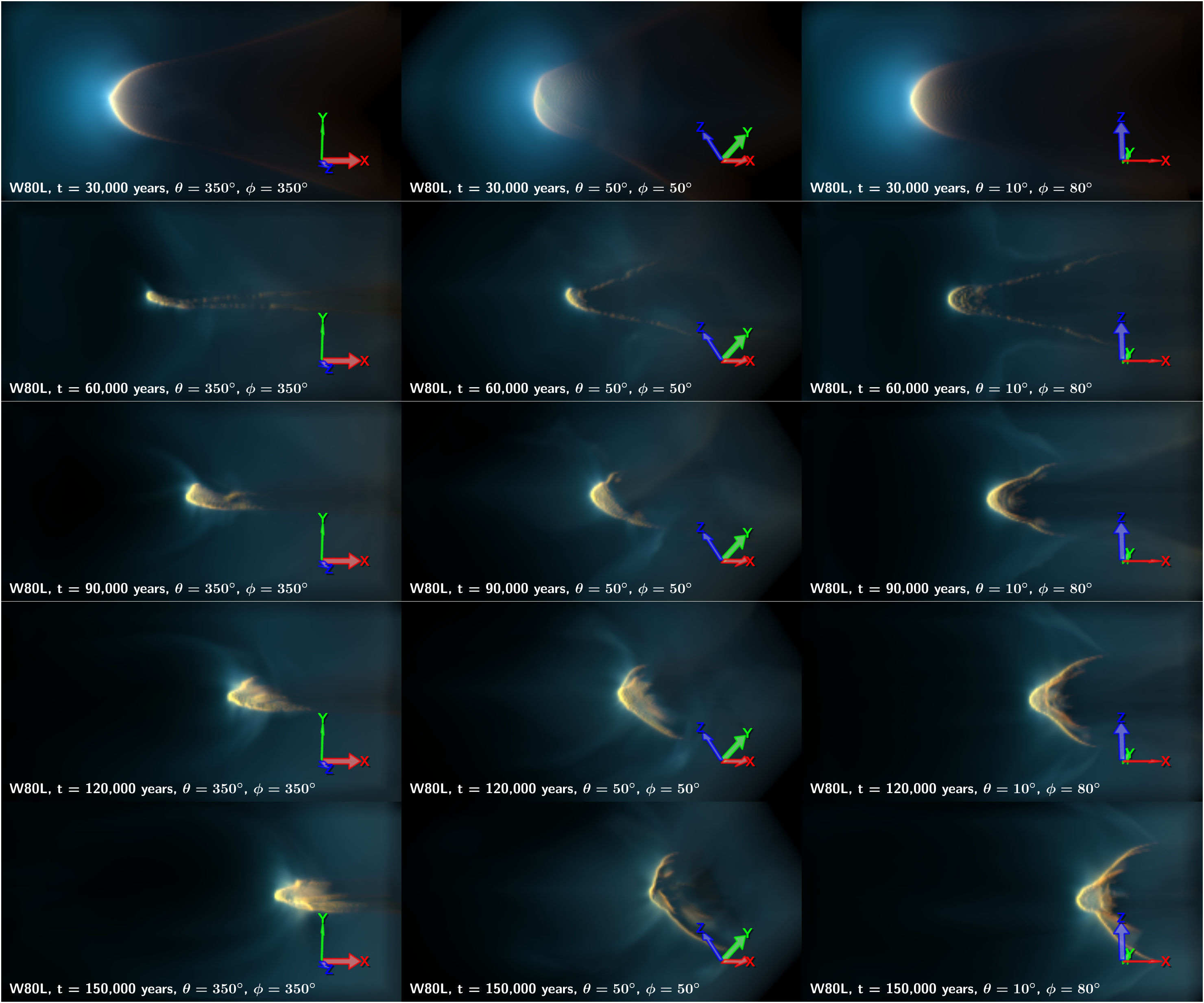}
  \caption{\New{As Fig.~\ref{fig:zero}, but showing multiple views of run W80L for a sequence of evolutionary times between 30,000 and 120,000~years. View directions are as in Fig.~\ref{fig:views}. }}
  \label{fig:W80L}
\end{figure*}
This run has an initial magnetic field that is \(\surd{}10\) times smaller than in the strong-field runs, giving an initial plasma parameter of \(\beta_0 = 0.1\). Although the magnetic pressure dominates the thermal pressure in the initial globule, the thermal pressure of the ionised photoevaporation flow is much greater than this and so the magnetic field plays a minor r\^ole in the subsequent evolution of the globule, which is more similar to the non-magnetic case (\S~\ref{sec:zero-field:-z00l}) than to the strong-field case (\S~\ref{sec:strong-perp-field}). In particular, the flattening of the globule along the field lines (\S~\ref{sec:early-evolution}) \New{is much less extreme with an aspect ratio of about 2:1. In addition}, no current sheet forms at the symmetry plane (\S~\ref{sec:early-evolution}), and the photoevaporation flow is powerful enough to drive all ambient material off the grid, so there is no recombining shocked shell (\S~\ref{sec:late-evolution:-s90l}). \New{The evolution of this case is shown in Figure~\ref{fig:W80L}.}

\subsection{Comparative study of global globule evolution}
\label{sec:comp-study-glob}

\begin{figure}
  \centering
  \includegraphics[width=\linewidth, trim=0 30 0 0, clip]{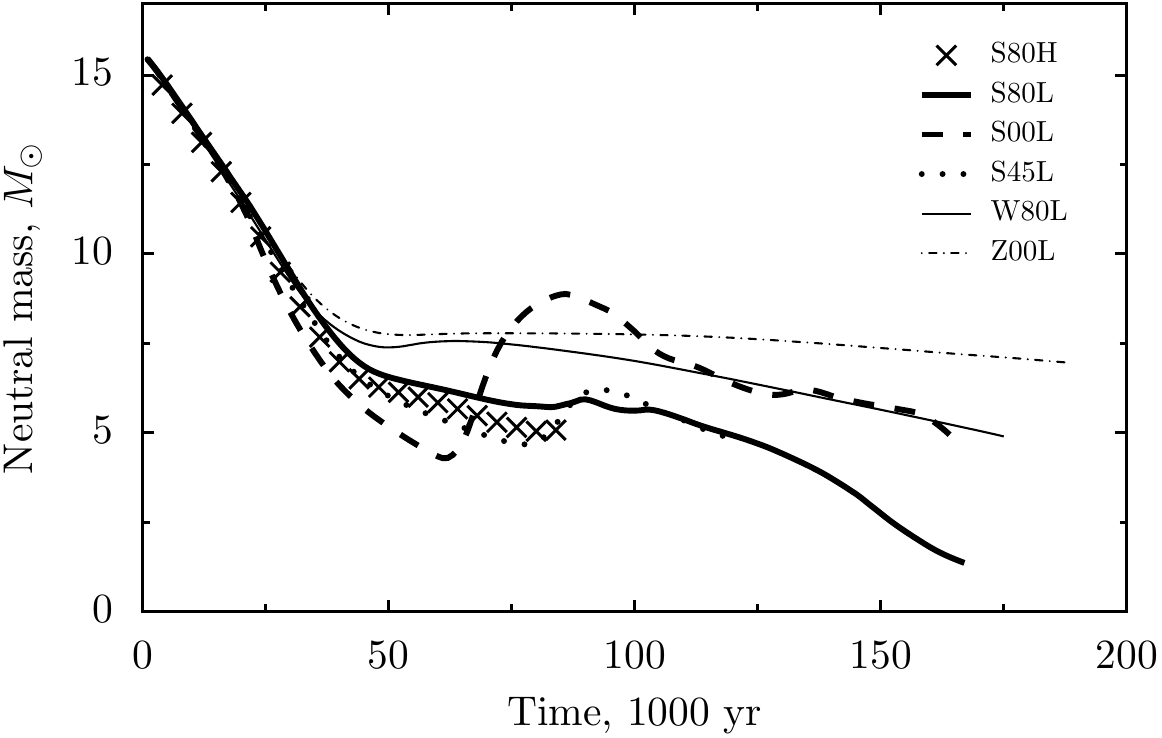}
  \includegraphics[width=\linewidth]{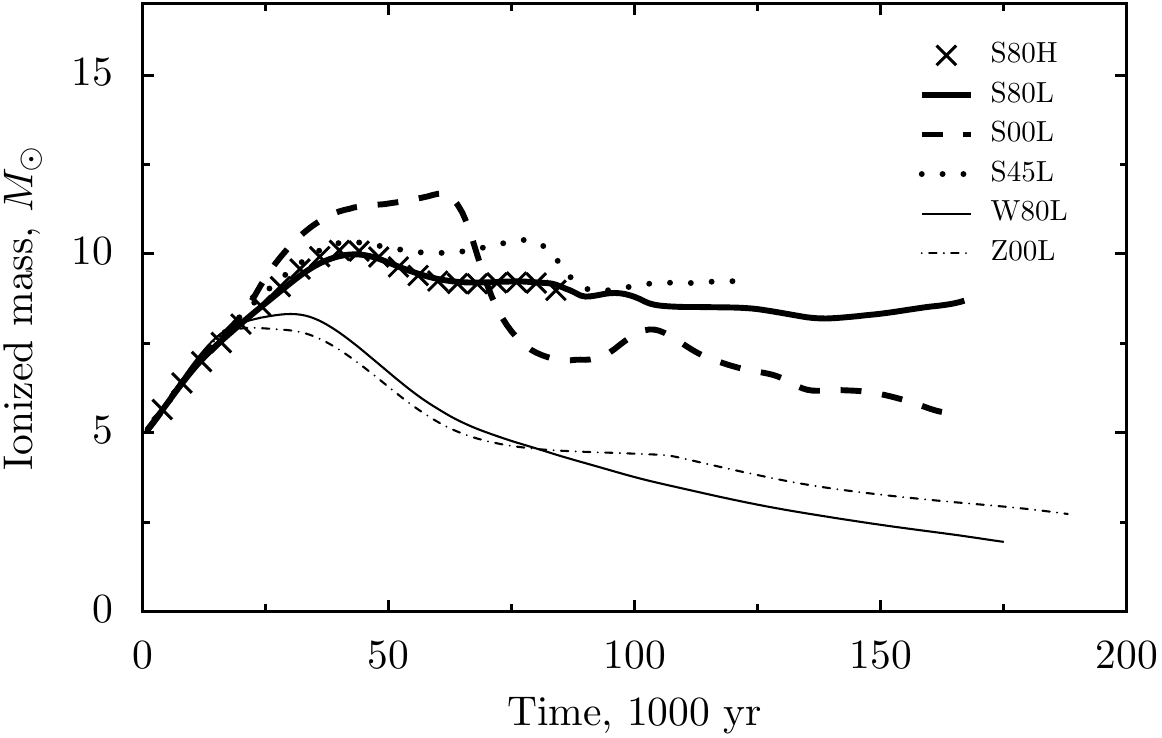}
  \caption{Evolution with time of the total mass of neutral gas (top)
    and ionised gas (bottom). Different line types and symbols show
    results for different model runs, as indicated in the key.}
  \label{fig:mass}
\end{figure}
\begin{figure}
  \centering
  \includegraphics[width=\linewidth, trim=0 30 0 0, clip]{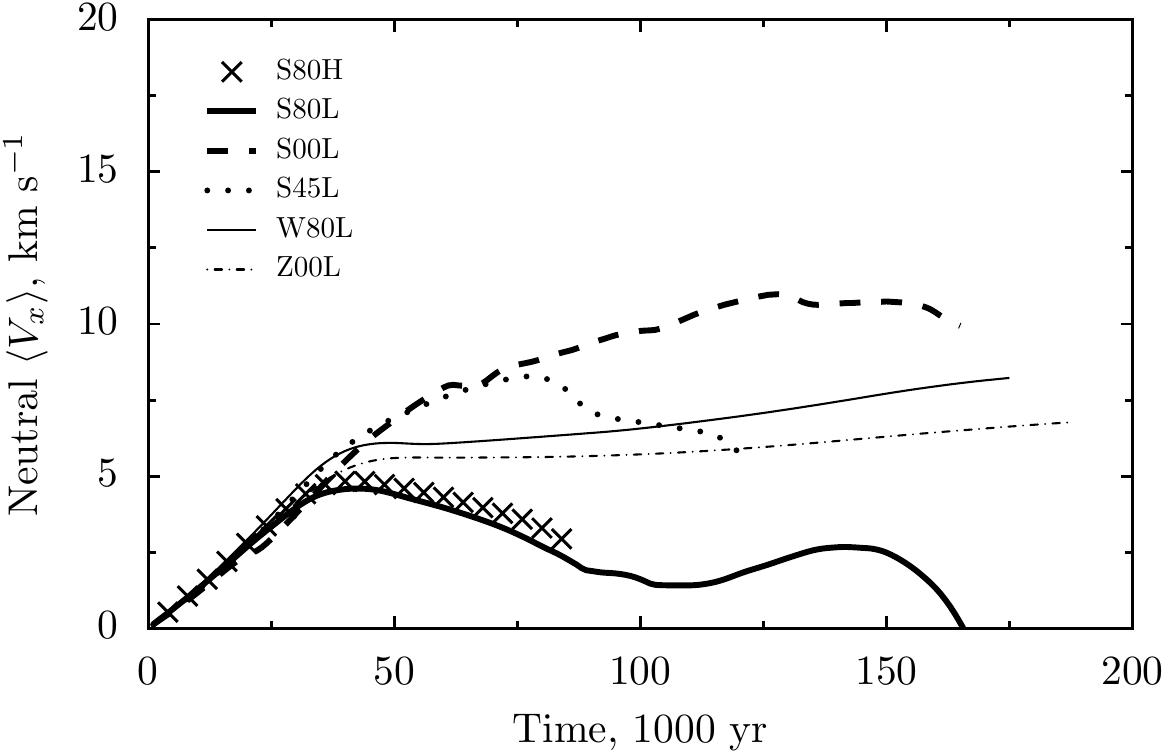}
  \includegraphics[width=\linewidth]{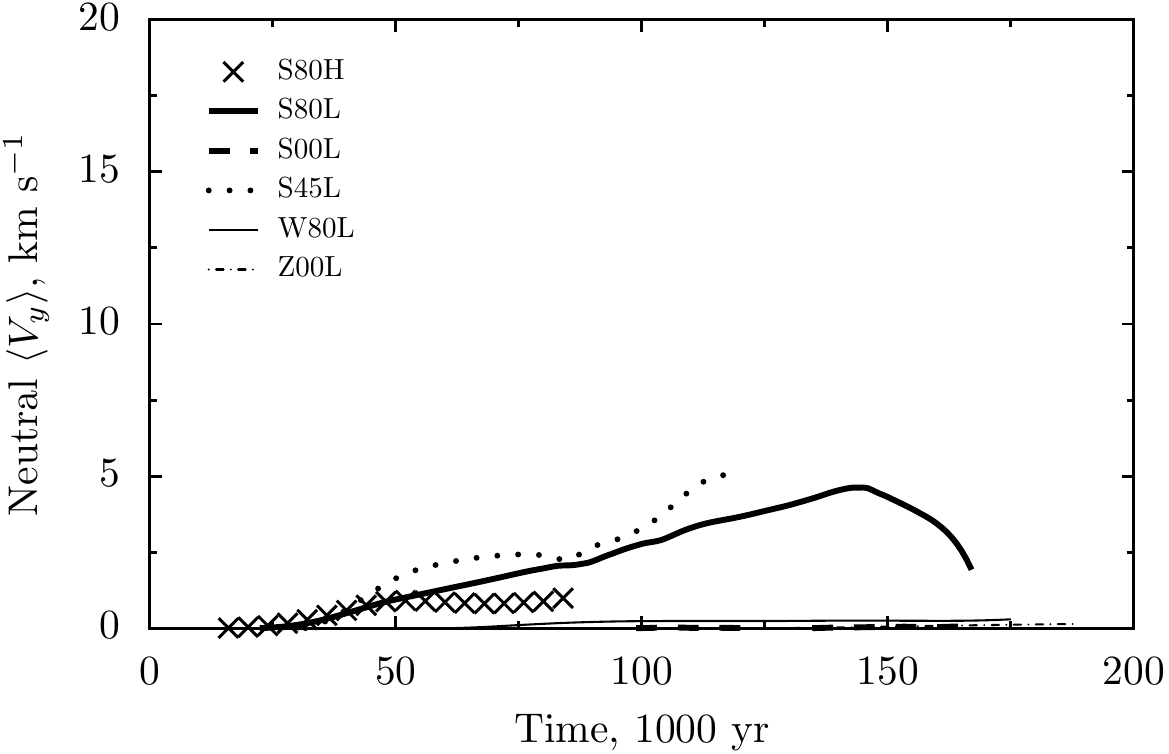}
  \caption{As Fig.~\ref{fig:mass}, but showing the evolution with time
    of the velocity of the centre of mass of the neutral gas. Top
    panel shows velocity component along the \(x\)-axis. Bottom panel
    shows velocity component along the \(y\)-axis.}
  \label{fig:meanv}
\end{figure}
\begin{figure}
  \centering
  \includegraphics[width=\linewidth , trim=0 30 0 0, clip]{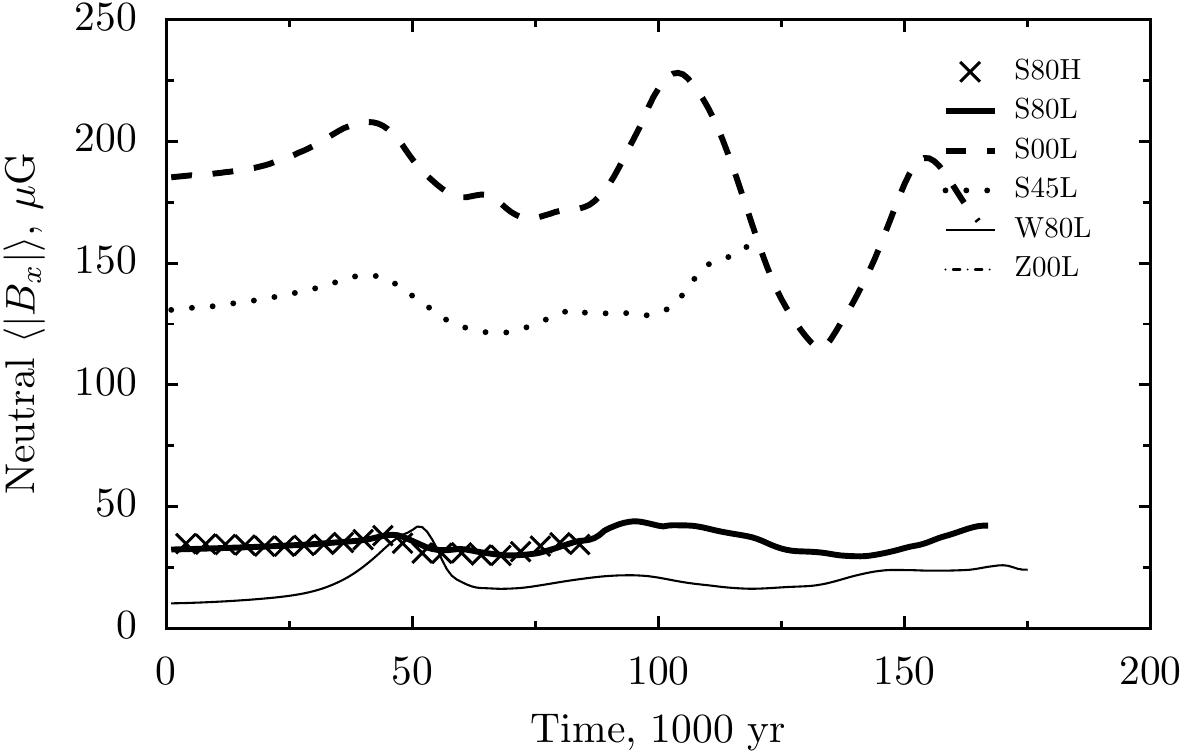}
  \includegraphics[width=\linewidth]{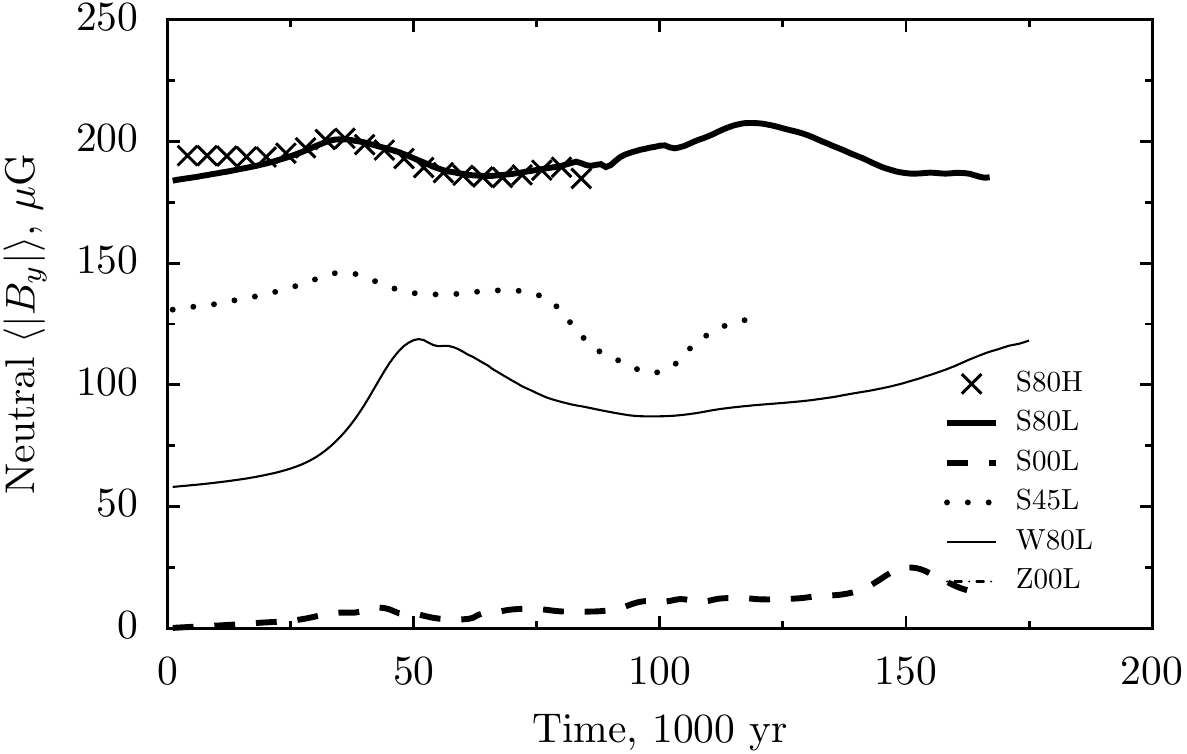}
  \caption{As Fig.~\ref{fig:mass}, but showing the evolution with time
    of the mean magnetic field in the neutral gas. Top panel shows
    \(x\)-component of field. Bottom panel shows \(y\)-component of
    field.}
  \label{fig:meanb}
\end{figure}
\begin{figure}
  \centering
  \includegraphics[width=\linewidth , trim=0 30 0 0, clip]{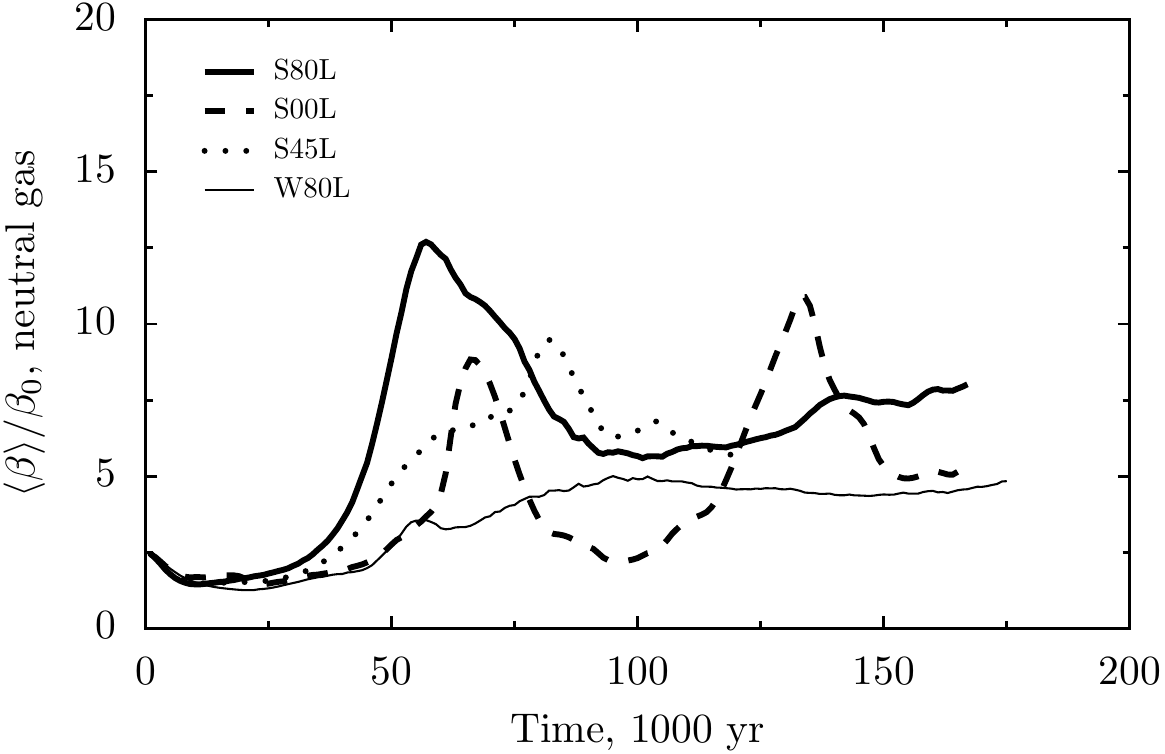}
  \includegraphics[width=\linewidth]{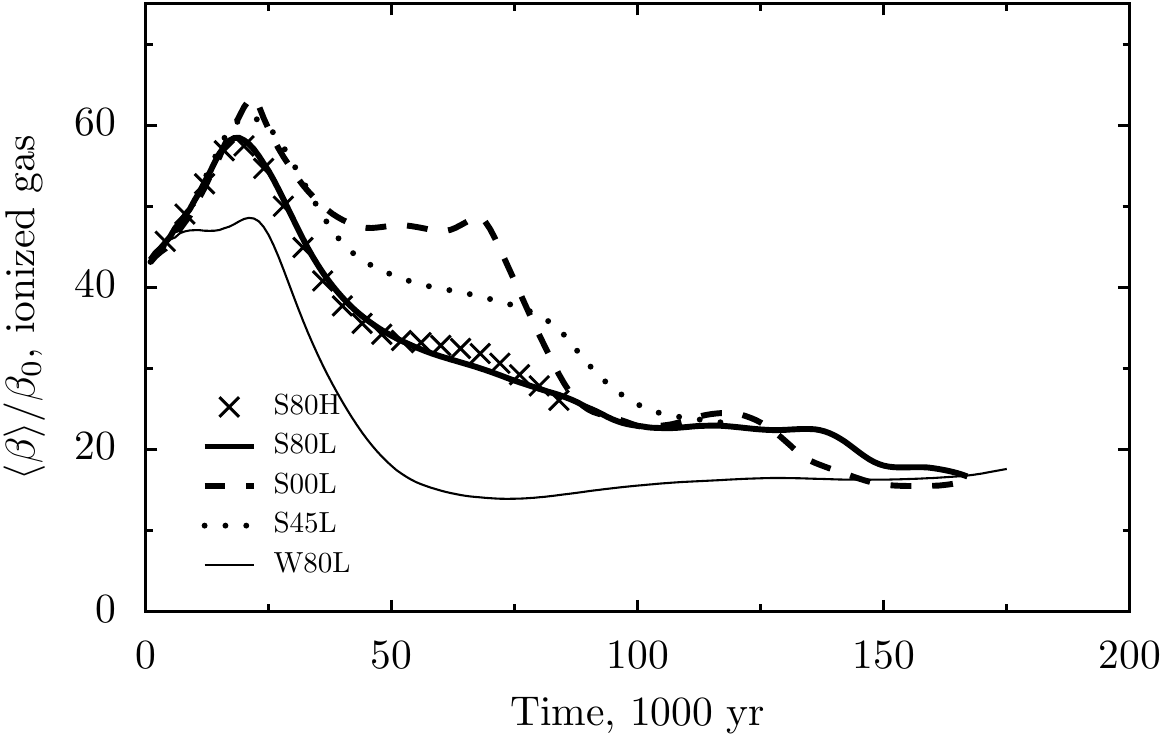}
  \caption{As Fig.~\ref{fig:mass}, but showing the evolution with time
    of the mean plasma parameter, scaled by the initial value
    \(\beta_0\), for neutral gas (top) and ionised gas (bottom). \New{Note the different vertical scales in the two graphs.}}
  \label{fig:meanbeta}
\end{figure}

Figures~\ref{fig:mass} to \ref{fig:meanbeta} compare the evolution of various global average properties between the different runs. For all the runs, the mass of neutral gas (Fig.~\ref{fig:mass}, top) falls in an identical manner during the initial stages of shock compression, as the globule begins to accelerate (Fig.~\ref{fig:meanv}), but after about 20,000~yr small differences become apparent. The strongly magnetised globules all evaporate faster than the weakly magnetised and unmagnetised globules, but for different reasons. In the \(\theta_0 = 0^\circ\) case (run S00L), where the effect is the strongest, it is because the lateral compression of the globule is retarded by the magnetic field (\S~\ref{sec:strong-parall-field}), so that the globule presents a larger cross-section to the ionising radiation, increasing the photoevaporation rate. In the \(\theta_0 = 80^\circ\) case (runs S80L and S80H), the evaporation rate does not begin to exceed the non-magnetic case until after 40,000~years, at which point the globule center-of-mass velocity has stopped increasing (Fig.~\ref{fig:meanv}, top) due to braking by the threaded field lines, so that at a given time the globule is closer to the star than in the non-magnetic case and is therefore exposed to a greater ionising flux.

An extended plateau is seen in the mass of ionised gas (Fig.~\ref{fig:mass}, bottom) after \(\simeq 30,000\)~years, during which time the globule is in its equilibrium cometary phase. The amount of ionised gas that is retained on the grid is much higher in the strongly magnetised runs, due to the trapping of the ionised photoevaporation flow in a shell (\S~\ref{sec:strong-perp-field}). Apart from in the \New{\(\theta_0 = 80^\circ\) strong field} case, where magnetic braking is efficient, all the runs show a very similar evolution in the globule center-of-mass velocity along the \(x\)-axis (Fig.~\ref{fig:meanv}, top). \New{In the slanted field runs (S45L, S80L)}, the globule develops a lateral velocity along the \(y\)-axis (Fig.~\ref{fig:meanv}, bottom). \New{Interestingly, the magnitude of the lateral velocity is similar in the two cases, although the time evolution is somewhat different. For \(\theta_0 = 45^\circ\) the globule moves approximately in a straight line with constant \(\langle V_y\rangle / \langle V_x \rangle\), whereas for \(\theta_0 = 80^\circ\),  \(\langle V_y\rangle / \langle V_x \rangle\) increases with time as the globule's path curves away from the radial direction.} 

\New{After a time between 60 and 80,000~years,} depending on the field inclination \(\theta_0\),  the equilibrium cometary phase for the strongly magnetised globules \New{is modified due to recombination} in the shocked ionised shell. \New{In the cases where this effect is greatest (S45L and S00L)}, this can be seen as an increase in the neutral mass (Fig.~\ref{fig:mass}, top), \New{which in the case of \(\theta_0 = 0^\circ\) shows multiple epsisodes of recombination and reionisation.}

The mean magnetic field for the ionised or neutral component is calculated as \(\langle \vert B_i \vert \rangle = \int \vert B_i \vert x_j\, dV / \int x_j\, dV\), where \(B_i\) is \(B_x\) or \(B_y\) and \(x_j\) is the ionised or neutral fraction of hydrogen, as appropriate. The evolution of the mean magnetic field in the neutral gas is shown in Figure~\ref{fig:meanb}, where it can be seen \New{that in the strong field cases the variations from the initial value  (Table~\ref{tab:models}) are rather slight (10--20\%)}.\footnote{In part, this is due to the fact that much of the contribution to the mean field comes from the globule tail, where the magnetic field is relatively unperturbed.} During the initial implosion phase, there is competition between the leading fast-mode shock, which amplifies \(B\), and the following slow-mode shock, which attenuates \(B\). The result is a \New{small} gradual increase in the mean \(B\). The first rebound after maximum compression reverses this trend, but the late-time evolution in the strong-field runs is complicated by the recombination of the ionised shell. \New{In the weak field case (W80L), the field amplification is more significant, reaching a factor of two.}

Figure~\ref{fig:meanbeta} shows the evolution of the mean plasma \(\beta\) parameter, \(\langle\beta\rangle = \langle P_\mathrm{gas} \rangle/ \langle P_\mathrm{mag} \rangle\) for both the neutral gas (top) and ionised gas (bottom), normalised by the initial value (\(\beta_0 = 0.1\) for the weak-field runs, \(\beta_0 = 0.01\) for the strong-field runs). In the ionised gas, \(\langle\beta\rangle/\beta_0\) \New{is initially in the range} 40--60, \New{falling to 10--20 at later stages}. Thus, on average, the magnetic pressure exceeds the gas pressure in the ionised region by a factor of \New{2--10}. However, this represents a compromise between the photoevaporation flow from the globule head, where the gas pressure dominates\footnote{Since the flow is supersonic, the gas pressure is in turn dominated by the ram pressure.} (\(\beta > 1\)), and the flanks of the tail, where the magnetic field dominates (\(\beta < 1\)). In the weak field case, the ionised \(\langle\beta\rangle/\beta_0\) falls off somewhat during the equilibrium cometary phase from the initial value of \(\simeq 40\), but the magnetic contribution to the total pressure in the ionised gas remains small.

In the neutral gas (Fig.~\ref{fig:meanbeta},top), \(\langle\beta\rangle/\beta_0\) \New{tends to increase} during the globule evolution, \New{but in the strong field cases the magnetic pressure remains dominant over the thermal pressure by about one order of magnitude.} During the initial implosion, \New{the increase in  \(\langle\beta\rangle\)} is due to pressurisation of the globule by shocks. The strong, almost perpendicular field run (S80L) shows a larger boost in the neutral \(\langle\beta\rangle\) than the other models, which is probably due in part to reconnection in the current sheet that forms in the globule midplane.

\New{Although} the evolution \New{of many} of the global properties in the strong perpendicular field case is almost indistinguishable between the low-resolution (S80L) and high-resolution (S80H) runs, \New{there are small but significant differences after the point of maximum compression (50,000~years), particularly in the neutral mass (Fig.~\ref{fig:mass} top) and the mean lateral velocity (Fig.~\ref{fig:meanv} bottom). This is probably due to the more vigorous thin-shell instabilities that are seen in the higher resolution run.}

% \TODO{Comparison with previous theoretical results. Bertoldi's
%   qualitative picture \citep{1989ApJ...346..735B}. Robin's 2d
%   simulations \citep{2007Ap&SS.307..179W}. Ryutov's funny models
%   \citep{2005Ap&SS.298..183R}.}

\New{
\subsection{Assessment of the numerical limitations of our simulations}
\label{sec:assessment}

\begin{figure*}
  \centering
  \includegraphics[trim=0 30 0 0, clip, scale=0.6] {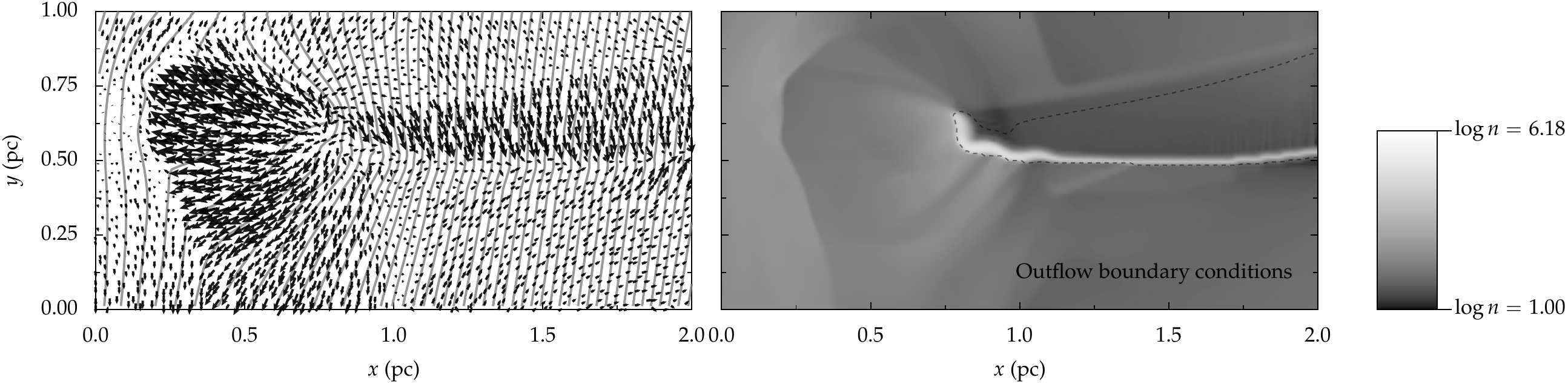}
  \includegraphics[trim=0 30 0 0, clip, scale=0.6] {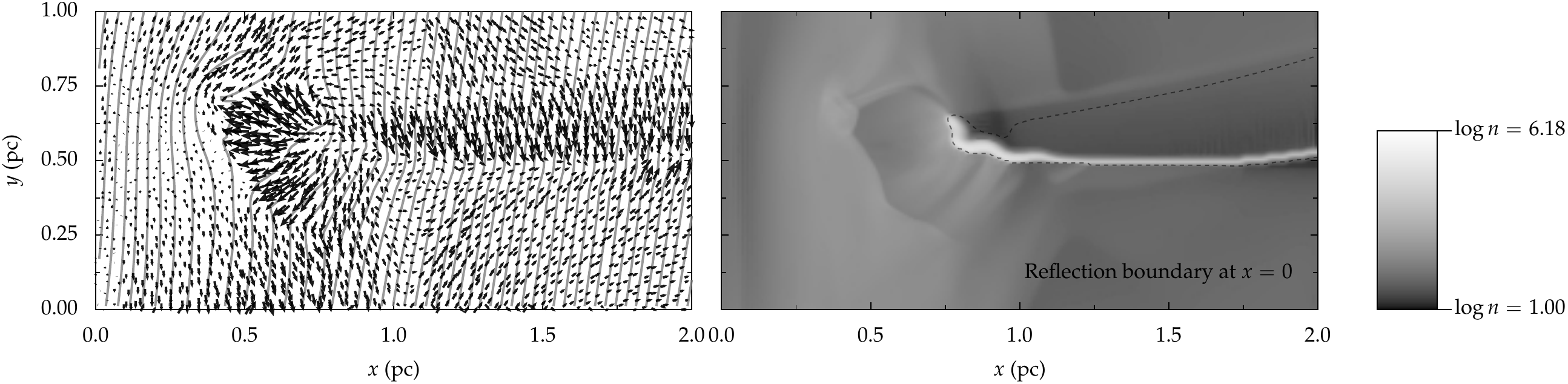}
  \includegraphics[trim=0 0 0 0, clip, scale=0.6] {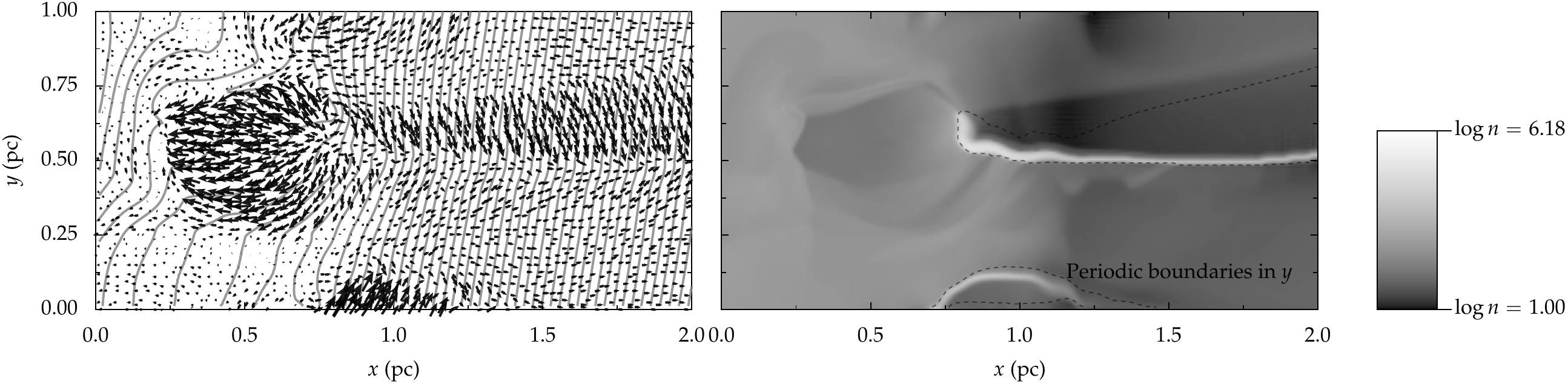}
  \caption[]{\New{As Fig.~\ref{fig:S80S} but showing the effects of boundary conditions on globule evolution. All simulations are similar to S80S but at a lower resolution of \(255 \times 127\) and each is shown at an evolutionary time of 80,000~years.}}
  \label{fig:bc}
\end{figure*}

The finite nature of the computational resources available to us means that our fixed-grid simulations have a rather modest number of grid points in each dimension (127 to 1002). This can potentially compromise the reliability of our results both on small scales and on large scales. At the smallest scales, we cannot resolve physical processes that occur on scales smaller than our cell size. To some extent, this is ameliorated by basing the governing equations on conservation laws and by careful treatment of the sub-grid physics, as in the special case of ionisation fronts \citep{2006NewA...11..374M}. However, other problems are less easily sidestepped, such as the grid viscosity and resistivity, which are many orders of magnitude larger than the true viscosity and resistivity of astrophysical plasmas. By comparing the behaviour of the same simulation carried out at different resolutions, we have determined that the lowest resolution that we employ (\(254 \times 127 \times 127\)) is broadly sufficient for modelling the global evolution of the globule, even though some of the finer scale behaviour is suppressed. In particular, the most compact configuration of the initial globule implosion is barely resolved in the low-resolution simulations and the subsequent fragmentation of the neutral globule is delayed and less vigorous. 

At the largest scales, the problem is that our computational domain is not many times larger than the size of the globule, so that the treatment of the boundaries of the simulation box may seriously affect the interior solution. The outflow boundary conditions that we employ (also called \emph{non-reflecting} or \emph{absorbing}, e.g., \citealp{Leveque:2002} \S~21.8.5) are unproblematic for cases where there is supersonic flow outward through the boundary, such as is found in the non-magnetic run Z00L\@.  However, in the strongly magnetised runs, the flow near the boundaries is frequently sub-Alfvenic and even subsonic, in which case no truly satisfactory treatment of the boundary conditions exists. We have attempted to quantify the effects that boundaries have on our simulations by carrying out various experiments in which the positions of the boundaries or the boundary conditions are varied. 

For instance, we have carried out an identical simulation to S80L, but with the globule-star system shifted 0.5~pc along the positive \(x\)-axis. This means that less of the tail is included on the grid but that the termination shock of the globule head's photoevaporation flow is moved well away from the \(x = 0\)  boundary. Some very minor differences are seen in the detailed structure of the shocked ribbon that forms at late times (\S~\ref{sec:late-evolution:-s90l}), but the timing of the recombination and the nature of the subsequent fragmentation are identical. 

In another series of experiments, we have carried out 2D simulations, similar to S80S, but at lower resolution, for three different sets of boundary conditions (Fig.~\ref{fig:bc}). In the control experiment (top panel) the boundary conditions on all faces are the normal outflow conditions.  In a second experiment (middle panel) we impose reflection boundary conditions at \(x = 0\), which is equivalent to there being two globules along the \(x\) axis, one on either side of the star at \((x, y) = (\pm 0.5, 0.0)\). In this case, the shock structures in the ionised photoevaporation flow from the head of the globule are considerably modified from the standard case, since the flow is now shocking against the mirror-image flow from the other side of the star. The magnetic field lines can no longer be pushed out of the computational domain through the \(x = 0\) boundary and the resultant increase in magnetic pressure means that the termination shock of the photoevaporation flow occurs closer to the globule than in the free-outflow case.

In a third experiment (bottom panel) we have imposed periodic boundary conditions at \(y = 0\) and \(y = y_{\mathrm{max}}\), which is equivalent to stacking a series of identical globules on top of one another. Again, the shock structures in the photoevaporation flow are modified since as well as shocking against the ambient medium in the \(-x\) direction, the flow also shocks against the equivalent flows from the ``other'' globules in the \(\pm y\) directions. Note that the formation of a dense neutral shell and shadow zone around \((x, y) = (1.0, 0.0)\) is an unphysical artefact of the fact that we have only employed periodic boundary conditions for the MHD and not for the radiative transfer. In reality, the shell should not form, since the shadow would be ionised by the ``other'' star that is located off the grid at \((x, y) = (0.0, -0.5)\). In this experiment, the accretion flow onto the tail is less dense than in the standard case, since there is no longer an infinite reservoir of material outside of the box that can be sucked in along field lines through the \(y\) boundaries. 

Note that despite the changes in the structure of the photoevaporation flow in the experiments with different boundary conditions, the evolution of the neutral globule remains hardly affected. The potential for  ``sucking in'' material through the boundaries is the least satisfactory property of the outflow boundary conditions (which could equally well be called inflow conditions). However, we have determined that the effects of this on our simulations are generally neglible. One exception is the late-time evolution of run S80S (Fig.~\ref{fig:S80S}), where, after \(t = 60,000\)~years,  a stream of inflowing material can be seen to develop from the \(y = 1.0\) boundary at \(x \simeq 1.75\), which causes distortions in the far tail after \(t = 100,000\)~years. 

The only reliable way of eliminating spurious boundary effects would be to make the computational domain so large that all physical quantities have become approximately uniform before the boundary is reached. However, this approach is computationally unfeasible with the uniform-grid algorithms employed in this paper. Furthermore, real photoevaporated globules do not exist in isolation, but are embedded in complex dynamic environments. Therefore, exploring a range of possible boundary conditions, as we have in this section, is the best way to determine to what extent the globule evolution is affected by its large-scale environment.
}

\section{Comparison with observations}
\label{sec:observ-magn-glob}\label{sec:discussion}

Magnetic fields have been measured or inferred by various direct and indirect methods, both for cometary globules, which show evidence of interaction with the radiation field of nearby massive stars, as well as for dark globules, which show no evidence of such an interaction. Measurements of both types of globule are relevant to our simulations since the dark globules represent the initial conditions before the photoevaporation process starts. 

Magnetic field measurements in dark globules are typically based on polarisation measurements of the far-infrared emission from dust grains \citep{2001ApJ...561..871H, 2003ApJ...588..910V, 2003ApJ...592..233W, 2007AJ....134..628V}, which probes dense gas that is very optically thick at visual wavelengths. The projected magnetic field direction on the plane of the sky can be determined from the orientation of the polarisation vectors, while the field strength can be estimated from the dispersion in polarisation direction by the Chandrasekhar-Fermi method, so long as the gas density and turbulent velocity dispersion are known. The inferred field strengths and densities range from 100 to 300~\(\mu\)G, and from \(10^5\) to \(10^6~\pcc\), respectively, in globules with total masses from 0.3 to \(100~M_\odot\) and sizes around 0.1~pc. 

The magnetic field in isolated cometary globules has been investigated in a series of papers \citep{1996MNRAS.279.1191S, 1999MNRAS.308...40B, 2004MNRAS.348...83B} via the visual-band polarisation, \(p_V\), of background stars. As with the far-infrared measurements, this directly reveals the projected field direction, and can be also used to estimate the field strength from the \(p_V/A_V\) ratio, where \(A_V\) is the visual-band extinction, if one assumes that the grain alignment efficiency is the same as in the general ISM\@. In contrast to the far-infrared measurements, this technique samples the lower density gas in the globule envelope. The only globule where the field strength has been measured by this technique is CG~22 in the Gum Nebula \citep{1996MNRAS.279.1191S}, giving a value of \(30~\mu\)G in gas with a mean density of \(200~\pcc\), together with some evidence that the field is stronger (\(70~\mu\)G) in the higher density (\( n > 1000~\pcc\)) core.  This globule, with a radius of  \(0.6\)~pc, is considerably larger than the dark  globules discussed above, but is comparable in mass (\(\sim 15~M_\odot\)). The field direction is parallel to the globule tail in this case, but in another globule in the same region (CG~30--31; \citealp{1999MNRAS.308...40B}) the field is found to be perpendicular to the tail.\footnote{In a third globule with a measured field direction, CG~12 \citep{2004MNRAS.348...83B}, the cometary shape is believed to be due interaction with a supernova remnant, so it is not directly relevant to  the current models.}

In summary, over changes of four orders of magnitude in globule density (\(\sim 100\) to \(\sim 10^6~\pcc\)), the linear polarization observations show only about one order of magnitude variation in \New{globule} magnetic field strengths (\(B \sim 40\) to \(\sim 400~\mu\)G, after correcting for the unobserved line-of-sight component). Assuming a similar gas temperature in all the globules, this implies considerable variation in the plasma \(\beta\) parameter (ratio of thermal pressure to magnetic pressure), ranging from \(\beta \sim 0.01\) for the larger, low-density globules to \(\beta \sim 1\) for the more compact, dense, dark globules. \New{However, this picture is rather different from the results of Zeeman measurements of the line-of-sight component of magnetic field in molecular cloud cores \citep{1999ApJ...520..706C, 2005LNP...664..137H, 2008A+A...487..247F}, which suggest a roughly constant value of \(\beta \sim 0.05\) up to the highest measured densities (\(\sim 10^7~\pcc\)). It is unclear whether this discrepancy is an artefact of the difference in observational technique or represents a real difference between the two classes of objects. The molecular cloud cores do tend to be much more massive (\(\sim 10\)--\(1000~M_\odot\)) than the isolated globules (\(\sim 0.1\)--\(10~M_\odot\)).}

% Although this is based on a rather small number of observations, it is consistent with a much larger set of measurements for the ISM as a whole \citep{2005LNP...664..137H}.

From our numerical results (\S~\ref{sec:results}), we find that \New{although deviations from cylindrical symmetry in the compressed globule occur for \(\beta_0 = 0.1\), the strongest effects of the magnetic field on the globule evolution require \(\beta_0 = 0.01\).} However, magnetically dominated globule evolution may be more common than this fact suggests, since classical, isolated cometary globules, such as those in the Gum Nebula \citep{1983A&A...117..183R}, are typically exposed to a weaker ionising radiation field than was employed in our simulations. Two-dimensional simulations of globules exposed to a weaker radiation field \citep{2007Ap&SS.307..179W} show that magnetic effects can be important even when \(\beta_0 \sim 1\) for the case where the ionisation front in the ambient medium is D-type when it passes the globule.

% \TODO{Geometry}

In addition to the clearly isolated globules discussed above, another class of objects that can be compared with our simulations is a whole range of structures seen inside or at the edges of \ion{H}{2} regions, which, according to their morphology, may be variously described as bright rims \citep{1956BAN....13...77P}, fingers/columns/pillars/elephant trunks  \citep[hereafter simply pillars]{1949PASP...61..151M,  1957ApJ...125..622O, 2004A&A...418..563R}, or bars \citep{2000AJ....120..382O}. Some of the diversity in terminology is spurious, with multiple terms being employed for identical or similar structures, sometimes within the same paper. Similar objects that have been observed at smaller size scales have given rise to yet further profusion of terms: evaporating gaseous globules, or EGGS \citep{1996AJ....111.2349H}, proplyd-like features \citep{2006AJ....131.2580D}, and globulettes \citep{2007AJ....133.1795G}. In all cases, the underlying physical process is probably the same: a dense condensation in the neutral gas slows down the propagation of the ionisation front, causing it to wrap around and further compress the obstacle. It is usually unclear in individual objects whether the dense neutral condensation was pre-existing \citep{2006ApJ...647..397M}, or whether it formed due to an instability in the propagation of the ionisation front \citep{1996ApJ...469..171G, 1999MNRAS.310..789W}. Numerical experiments \citep{2001MNRAS.327..788W} show that the resultant structure and evolution of the photoevaporated condensation are remarkably similar in the two cases. 

There is one sub-class of these objects that seems to show a real physical difference from the others: the linear, bright emission features sometimes referred to as bars, of which the Orion Bright Bar is the prototypical example \citep[and references therein]{2001ARA&A..39...99O}. Unlike the majority of globules and pillars, these features have a long axis that is oriented approximately perpendicular (in projection) to the direction of the incident ionising radiation. This is rather reminiscent of the emission structures produced in our simulations S80H and S80L (Figures~\ref{fig:views} and \ref{fig:ribbon}), suggesting that a strong perpendicular magnetic field may play a role in the formation of bar-like features. Magnetic field measurements in the Orion Bar \citep{1990ApJ...357..132G, 2004ApJ...604..717H, Kusakabe:2008} are broadly consistent with this scenario: unlike in the main Orion molecular filament, where the field orientation is close to the plane of the sky and projected along the filament length, the field in the Bar is at an angle\footnote{Note that even if the magnetic field is perpendicular to the Bar's axis, this will not generally be true in projection.} to the projected long-axis of the Bar, with indirect evidence that its orientation is closer to the line of sight. However, the true three-dimensional orentations of the Bar, magnetic field, and radiation field are far from clear and require further study in order to establish whether or not the Orion Bar has been shaped by the effects of magnetically modified photoevaporation and implosion.

The initial conditions employed in the globule simulations reported in this paper are highly idealised. In particular, the assumption of a uniform magnetic field threading the entire simulation volume is unlikely to be satisfied in real objects. In a companion paper \citep{Arthur:2009}, we report on global simulations of \ion{H}{2} region evolution in highly inhomogeneous, turbulent, magnetised media. In those simulations the formation of globule-like photoevaporation flows is an essential feature, \New{as has previously been studied in the non-magnetic case \citep{2006ApJ...647..397M, 2007ApJ...668..980M}.}
 
%\TODO{ In reality, field strength should go up with density (Alfv\'en speed roughly constant for the higher density ranges. Difficulty in constructing a divergence-free equilibrium configuration for the initial conditions if field is not uniform.)}

%\TODO{Compare the direct effect of radiation pressure on the acceleration of the globule. Let's hope this is small, since we don't include it.}

% \TODO{Talk about self-gravity. Evolution of mass-to-flux ratio? Fact that evaporation time is much less than free-fall time. }

% \TODO{Put in context of general \ion{H}{2} region evolution. Refer to companion paper \citep{Arthur:2009}.}

\section{Conclusions}
\label{sec:conclusions}

We have carried out the first three-dimensional radiation-magneto\-hydro\-dynamic simulations of the evolution of a dense, magnetised globule that is exposed to a source of ionising radiation. For the globule parameters that we employ, \New{we find that when the initial magnetic field in the globule is strong (magnetic pressure \(\ga 100\) times gas pressure) there are} significant effects on the globule evolution, \New{such as the following:}
\begin{enumerate}
\item Strong deviations from cylindrical symmetry in the shape of the radiatively imploded globule. These deviations are greatest when the initial magnetic field orientation is \New{close to} perpendicular to the direction of the ionising radiation, in which case the shocked globule adopts a flattened, plate-like form. In the case of \New{a more} obliquely oriented initial field, the shocked globule ``curls up'' to form a comma-shaped structure.  
\item Modification of the rocket effect, which, in the non-magnetic case, accelerates the globule away from the ionising source to reach velocities of \(> 10~\kms\). For a \New{close to} perpendicular field orientation, the globule acceleration is reversed at late times, whereas for an oblique field orientation the globule trajectory is merely deflected from a purely radial motion.
\item Magnetic confinement of the ionised photoevaporation flow from the globule head, which eventually leads to trapping of the ionisation front in a shocked shell, and the temporary recombination of \New{parts} of the ionised flow. The subsequent evolution is complex, but may lead to fragmentation of the shocked shell and the formation of \New{secondary} mini-globules. 
\end{enumerate}
In the case where the initial magnetic field direction is exactly parallel to the direction of the ionising radiation, only the last of these three effects applies. In all strong-field cases that we have considered, the ionisation front jump conditions on the head of the globule are generally of the ``extra strong'' type, in which the magnetic field becomes aligned perpendicular to the ionisation front. The densest parts of the ionised photoevaporation flows are always found to be gas-pressure dominated, \New{although the ionised gas as a whole is magnetically dominated.} 

When the magnetic field is weaker (magnetic pressure \(\la 10\) times gas pressure in the initial globule), then the globule evolution is \New{more} similar to the non-magnetic case, \New{with the principal difference being a moderate flattening of the compressed globule.} 

We find evidence that magnetic effects may be important in the formation of bright, bar-like emission features in \ion{H}{2} regions.

\New{Our simulations include a careful treatment of the heating and cooling in the neutral/molecular gas and we find typical temperatures of \(50\) to \(100\)~K for the densest material, maintained by a combination of shocks and x-ray heating. This is warmer than has been assumed in previous studies, suggesting that gravitational instability of the imploded globule may be less important than has been commonly inferred.}

\section*{Acknowledgements}
\label{sec:acknowledgements-}

Part of the numerical simulations reported in this paper were carried out at the Departamento de Superc\'omputo, Direcci\'on General de Servicios de C\'omputo Acad\'emico, Universidad Nacional Aut\'onoma de M\'exico. WJH and SJA are grateful for financial support from DGAPA-UNAM, Mexico (PAPIIT IN112006, IN110108 and IN100309).  FDC acknowledges support by the European Community's ``Marie Curie Actions -- Human Resource and Mobility'' within the JETSET (Jet Simulations, Experiments and Theory) network under contract MRTN-CT-2000~005592. \New{We are grateful to an anonymous referee for perceptive comments that have led to significant improvements in the paper.}

\bibliographystyle{mn2e}
\bibliography{BibdeskLibrary}

\begin{thebibliography}{}

\bibitem[\protect\citeauthoryear{{Abel}, {Hoof}, {Shaw}, {Ferland} \&
  {Elwert}}{{Abel} et~al.}{2008}]{Abel:2008}
{Abel} N.~P.,  {Hoof} P.~A.~M.~v.,  {Shaw} G.,  {Ferland} G.~J.,    {Elwert}
  T.,  2008, \apj, 686, 1125

\bibitem[\protect\citeauthoryear{{Arthur}, {Henney}, {Mellema}, {De Colle} \&
  {V{\'a}zquez-Semadeni}}{{Arthur} et~al.}{2009}]{Arthur:2009}
{Arthur} S.~J.,  {Henney} W.~J.,  {Mellema} G.,  {De Colle} F.,
  {V{\'a}zquez-Semadeni} E.,  2009, \apj, in preparation

\bibitem[\protect\citeauthoryear{{Axford}}{{Axford}}{1964}]{1964ApJ...140..112%
A}
{Axford} W.~I.,  1964, \apj, 140, 112

\bibitem[\protect\citeauthoryear{{Baldwin}, {Ferland}, {Martin}, {Corbin},
  {Cota}, {Peterson} \& {Slettebak}}{{Baldwin}
  et~al.}{1991}]{1991ApJ...374..580B}
{Baldwin} J.~A.,  {Ferland} G.~J.,  {Martin} P.~G.,  {Corbin} M.~R.,  {Cota}
  S.~A.,  {Peterson} B.~M.,    {Slettebak} A.,  1991, \apj, 374, 580

\bibitem[\protect\citeauthoryear{{Bedijn} \& {Tenorio-Tagle}}{{Bedijn} \&
  {Tenorio-Tagle}}{1984}]{1984A&A...135...81B}
{Bedijn} P.~J.,  {Tenorio-Tagle} G.,  1984, \aap, 135, 81

\bibitem[\protect\citeauthoryear{{Bertoldi}}{{Bertoldi}}{1989}]{1989ApJ...346.%
.735B}
{Bertoldi} F.,  1989, \apj, 346, 735

\bibitem[\protect\citeauthoryear{{Bertoldi} \& {Draine}}{{Bertoldi} \&
  {Draine}}{1996}]{1996ApJ...458..222B}
{Bertoldi} F.,  {Draine} B.~T.,  1996, \apj, 458, 222

\bibitem[\protect\citeauthoryear{{Bertoldi} \& {McKee}}{{Bertoldi} \&
  {McKee}}{1990}]{1990ApJ...354..529B}
{Bertoldi} F.,  {McKee} C.~F.,  1990, \apj, 354, 529

\bibitem[\protect\citeauthoryear{{Bhatt}}{{Bhatt}}{1999}]{1999MNRAS.308...40B}
{Bhatt} H.~C.,  1999, \mnras, 308, 40

\bibitem[\protect\citeauthoryear{{Bhatt}, {Maheswar} \& {Manoj}}{{Bhatt}
  et~al.}{2004}]{2004MNRAS.348...83B}
{Bhatt} H.~C.,  {Maheswar} G.,    {Manoj} P.,  2004, \mnras, 348, 83

\bibitem[\protect\citeauthoryear{{Biro}, {Raga} \& {Canto}}{{Biro}
  et~al.}{1995}]{Biro:1995}
{Biro} S.,  {Raga} A.~C.,    {Canto} J.,  1995, \mnras, 275, 557

\bibitem[\protect\citeauthoryear{{Cant{\'o}}, {Raga}, {Steffen} \&
  {Shapiro}}{{Cant{\'o}} et~al.}{1998}]{1998ApJ...502..695C}
{Cant{\'o}} J.,  {Raga} A.,  {Steffen} W.,    {Shapiro} P.,  1998, \apj, 502,
  695

\bibitem[\protect\citeauthoryear{{Carlqvist}, {Gahm} \& {Kristen}}{{Carlqvist}
  et~al.}{2003}]{2003A&A...403..399C}
{Carlqvist} P.,  {Gahm} G.~F.,    {Kristen} H.,  2003, \aap, 403, 399

\bibitem[\protect\citeauthoryear{{Carlqvist}, {Kristen} \& {Gahm}}{{Carlqvist}
  et~al.}{1998}]{1998A&A...332L...5C}
{Carlqvist} P.,  {Kristen} H.,    {Gahm} G.~F.,  1998, \aap, 332, L5

\bibitem[\protect\citeauthoryear{{Cerqueira}, {Cant{\'o}}, {Raga} \&
  {Vasconcelos}}{{Cerqueira} et~al.}{2006}]{2006RMxAA..42..203C}
{Cerqueira} A.~H.,  {Cant{\'o}} J.,  {Raga} A.~C.,    {Vasconcelos} M.~J.,
  2006, Revista Mexicana de Astronomia y Astrofisica, 42, 203

\bibitem[\protect\citeauthoryear{{Colella} \& {Woodward}}{{Colella} \&
  {Woodward}}{1984}]{1984JCoPh..54..174C}
{Colella} P.,  {Woodward} P.~R.,  1984, Journal of Computational Physics, 54,
  174

\bibitem[\protect\citeauthoryear{{Crutcher}}{{Crutcher}}{1999}]{1999ApJ...520.%
.706C}
{Crutcher} R.~M.,  1999, \apj, 520, 706

\bibitem[\protect\citeauthoryear{{Dalgarno} \& {McCray}}{{Dalgarno} \&
  {McCray}}{1972}]{1972ARA&A..10..375D}
{Dalgarno} A.,  {McCray} R.~A.,  1972, \araa, 10, 375

\bibitem[\protect\citeauthoryear{{De Colle}}{{De Colle}}{2005}]{De-Colle:2005}
{De Colle} F.,  2005, PhD thesis, UNAM

\bibitem[\protect\citeauthoryear{{De Colle} \& {Raga}}{{De Colle} \&
  {Raga}}{2004}]{2004Ap&SS.293..173D}
{De Colle} F.,  {Raga} A.,  2004, \apss, 293, 173

\bibitem[\protect\citeauthoryear{{De Colle} \& {Raga}}{{De Colle} \&
  {Raga}}{2005}]{2005MNRAS.359..164D}
{De Colle} F.,  {Raga} A.~C.,  2005, \mnras, 359, 164

\bibitem[\protect\citeauthoryear{{De Colle} \& {Raga}}{{De Colle} \&
  {Raga}}{2006}]{2006A&A...449.1061D}
{De Colle} F.,  {Raga} A.~C.,  2006, \aap, 449, 1061

\bibitem[\protect\citeauthoryear{{De Marco}, {O'Dell}, {Gelfond}, {Rubin} \&
  {Glover}}{{De Marco} et~al.}{2006}]{2006AJ....131.2580D}
{De Marco} O.,  {O'Dell} C.~R.,  {Gelfond} P.,  {Rubin} R.~H.,    {Glover}
  S.~C.~O.,  2006, \aj, 131, 2580

\bibitem[\protect\citeauthoryear{{Dyson}}{{Dyson}}{1968}]{1968Ap&SS...1..388D}
{Dyson} J.~E.,  1968, \apss, 1, 388

\bibitem[\protect\citeauthoryear{{Dyson} \& {Williams}}{{Dyson} \&
  {Williams}}{1997}]{1997pism.book.....D}
{Dyson} J.~E.,  {Williams} D.~A.,  1997, {The physics of the interstellar
  medium}, 2nd edn.
Graduate series in astronomy, Bristol: Institute of Physics Publishing

\bibitem[\protect\citeauthoryear{{Esquivel} \& {Raga}}{{Esquivel} \&
  {Raga}}{2007}]{2007MNRAS.377..383E}
{Esquivel} A.,  {Raga} A.~C.,  2007, \mnras, 377, 383

\bibitem[\protect\citeauthoryear{{Falgarone}, {Troland}, {Crutcher} \&
  {Paubert}}{{Falgarone} et~al.}{2008}]{2008A+A...487..247F}
{Falgarone} E.,  {Troland} T.~H.,  {Crutcher} R.~M.,    {Paubert} G.,  2008,
  \aap, 487, 247

\bibitem[\protect\citeauthoryear{{Feigelson}, {Getman}, {Townsley}, {Garmire},
  {Preibisch}, {Grosso}, {Montmerle}, {Muench} \& {McCaughrean}}{{Feigelson}
  et~al.}{2005}]{2005ApJS..160..379F}
{Feigelson} E.~D.,  {Getman} K.,  {Townsley} L.,  {Garmire} G.,  {Preibisch}
  T.,  {Grosso} N.,  {Montmerle} T.,  {Muench} A.,    {McCaughrean} M.,  2005,
  \apjs, 160, 379

\bibitem[\protect\citeauthoryear{{Ferland}, {Korista}, {Verner}, {Ferguson},
  {Kingdon} \& {Verner}}{{Ferland} et~al.}{1998}]{1998PASP..110..761F}
{Ferland} G.~J.,  {Korista} K.~T.,  {Verner} D.~A.,  {Ferguson} J.~W.,
  {Kingdon} J.~B.,    {Verner} E.~M.,  1998, \pasp, 110, 761

\bibitem[\protect\citeauthoryear{{Flaccomio}, {Damiani}, {Micela}, {Sciortino},
  {Harnden} Jr., {Murray} \& {Wolk}}{{Flaccomio}
  et~al.}{2003}]{2003ApJ...582..382F}
{Flaccomio} E.,  {Damiani} F.,  {Micela} G.,  {Sciortino} S.,  {Harnden} Jr.
  F.~R.,  {Murray} S.~S.,    {Wolk} S.~J.,  2003, \apj, 582, 382

\bibitem[\protect\citeauthoryear{{Gahm}, {Grenman}, {Fredriksson} \&
  {Kristen}}{{Gahm} et~al.}{2007}]{2007AJ....133.1795G}
{Gahm} G.~F.,  {Grenman} T.,  {Fredriksson} S.,    {Kristen} H.,  2007, \aj,
  133, 1795

\bibitem[\protect\citeauthoryear{{Garc{\'\i}a-Segura} \&
  {Franco}}{{Garc{\'\i}a-Segura} \& {Franco}}{1996}]{1996ApJ...469..171G}
{Garc{\'\i}a-Segura} G.,  {Franco} J.,  1996, \apj, 469, 171

\bibitem[\protect\citeauthoryear{{Giuliani}
  Jr.}{{Giuliani}}{1979}]{1979ApJ...233..280G}
{Giuliani} Jr. J.~L.,  1979, \apj, 233, 280

\bibitem[\protect\citeauthoryear{{Gonatas}, {Engargiola}, {Hildebrand},
  {Platt}, {Wu}, {Davidson}, {Novak}, {Aitken} \& {Smith}}{{Gonatas}
  et~al.}{1990}]{1990ApJ...357..132G}
{Gonatas} D.~P.,  {Engargiola} G.~A.,  {Hildebrand} R.~H.,  {Platt} S.~R.,
  {Wu} X.~D.,  {Davidson} J.~A.,  {Novak} G.,  {Aitken} D.~K.,    {Smith} C.,
  1990, \apj, 357, 132

\bibitem[\protect\citeauthoryear{{Gonz{\'a}lez}, {Raga} \&
  {Steffen}}{{Gonz{\'a}lez} et~al.}{2005}]{2005RMxAA..41..443G}
{Gonz{\'a}lez} R.~F.,  {Raga} A.~C.,    {Steffen} W.,  2005, Revista Mexicana
  de Astronomia y Astrofisica, 41, 443

\bibitem[\protect\citeauthoryear{{Heiles} \& {Crutcher}}{{Heiles} \&
  {Crutcher}}{2005}]{2005LNP...664..137H}
{Heiles} C.,  {Crutcher} R.,  2005, in {Wielebinski} R.,  {Beck} R.,  eds,
  Cosmic Magnetic Fields Vol.~664 of Lecture Notes in Physics, Berlin Springer
  Verlag, {Magnetic Fields in Diffuse HI and Molecular Clouds}.
pp 137--+

\bibitem[\protect\citeauthoryear{{Henney}}{{Henney}}{2007}]{2007dmsf.book..103%
H}
{Henney} W.~J.,  2007, in {{Hartquist}, T.~W., {Pittard}, J.~M., \& {Falle},
  S.~A.~E.~G.} ed., , Diffuse Matter from Star Forming Regions to Active
  Galaxies.
Dordrecht: Springer, p.~103

\bibitem[\protect\citeauthoryear{{Henney}, {Arthur}, {Williams} \&
  {Ferland}}{{Henney} et~al.}{2005}]{2005ApJ...621..328H}
{Henney} W.~J.,  {Arthur} S.~J.,  {Williams} R.~J.~R.,    {Ferland} G.~J.,
  2005, \apj, 621, 328

\bibitem[\protect\citeauthoryear{{Henney}, {Williams}, {Ferland}, {Shaw} \&
  {O'Dell}}{{Henney} et~al.}{2007}]{2007ApJ...671L.137H}
{Henney} W.~J.,  {Williams} R.~J.~R.,  {Ferland} G.~J.,  {Shaw} G.,    {O'Dell}
  C.~R.,  2007, \apjl, 671, L137

\bibitem[\protect\citeauthoryear{{Henning}, {Wolf}, {Launhardt} \&
  {Waters}}{{Henning} et~al.}{2001}]{2001ApJ...561..871H}
{Henning} T.,  {Wolf} S.,  {Launhardt} R.,    {Waters} R.,  2001, \apj, 561,
  871

\bibitem[\protect\citeauthoryear{{Hester}, {Scowen}, {Sankrit}, {Lauer} \& {19
  others}}{{Hester} et~al.}{1996}]{1996AJ....111.2349H}
{Hester} J.~J.,  {Scowen} P.~A.,  {Sankrit} R.,  {Lauer} T.~R.,    {19 others}
  1996, \aj, 111, 2349

\bibitem[\protect\citeauthoryear{{Houde}, {Dowell}, {Hildebrand}, {Dotson},
  {Vaillancourt}, {Phillips}, {Peng} \& {Bastien}}{{Houde}
  et~al.}{2004}]{2004ApJ...604..717H}
{Houde} M.,  {Dowell} C.~D.,  {Hildebrand} R.~H.,  {Dotson} J.~L.,
  {Vaillancourt} J.~E.,  {Phillips} T.~G.,  {Peng} R.,    {Bastien} P.,  2004,
  \apj, 604, 717

\bibitem[\protect\citeauthoryear{{Hummer}}{{Hummer}}{1994}]{1994MNRAS.268..109%
H}
{Hummer} D.~G.,  1994, \mnras, 268, 109

\bibitem[\protect\citeauthoryear{{Iliev}, {Ciardi}, {Alvarez}, {Maselli},
  {Ferrara}, {Gnedin}, {Mellema}, {Nakamoto}, {Norman}, {Razoumov},
  {Rijkhorst}, {Ritzerveld}, {Shapiro}, {Susa}, {Umemura} \& {Whalen}}{{Iliev}
  et~al.}{2006}]{2006MNRAS.371.1057I}
{Iliev} I.~T.,  {Ciardi} B.,  {Alvarez} M.~A.,  {Maselli} A.,  {Ferrara} A.,
  {Gnedin} N.~Y.,  {Mellema} G.,  {Nakamoto} T.,  {Norman} M.~L.,  {Razoumov}
  A.~O.,  {Rijkhorst} E.-J.,  {Ritzerveld} J.,  {Shapiro} P.~R.,  {Susa} H.,
  {Umemura} M.,    {Whalen} D.~J.,  2006, \mnras, 371, 1057

\bibitem[\protect\citeauthoryear{{Kahn}}{{Kahn}}{1969}]{1969Phy....41..172K}
{Kahn} F.~D.,  1969, Physica, 41, 172

\bibitem[\protect\citeauthoryear{{Kessel-Deynet} \& {Burkert}}{{Kessel-Deynet}
  \& {Burkert}}{2003}]{2003MNRAS.338..545K}
{Kessel-Deynet} O.,  {Burkert} A.,  2003, \mnras, 338, 545

\bibitem[\protect\citeauthoryear{{Koyama} \& {Inutsuka}}{{Koyama} \&
  {Inutsuka}}{2002}]{2002ApJ...564L..97K}
{Koyama} H.,  {Inutsuka} S.-i.,  2002, \apjl, 564, L97

\bibitem[\protect\citeauthoryear{{Krumholz}, {Stone} \& {Gardiner}}{{Krumholz}
  et~al.}{2007}]{2007ApJ...671..518K}
{Krumholz} M.~R.,  {Stone} J.~M.,    {Gardiner} T.~A.,  2007, \apj, 671, 518

\bibitem[\protect\citeauthoryear{{Kusakabe}, {Tamura}, {Kandori}, {Hashimoto},
  {Nakajima}, {Nagata}, {Nagayama}, {Hough} \& {Lucas}}{{Kusakabe}
  et~al.}{2008}]{Kusakabe:2008}
{Kusakabe} N.,  {Tamura} M.,  {Kandori} R.,  {Hashimoto} J.,  {Nakajima} Y.,
  {Nagata} T.,  {Nagayama} T.,  {Hough} J.,    {Lucas} P.,  2008, \aj, 136, 621

\bibitem[\protect\citeauthoryear{{Lefloch} \& {Lazareff}}{{Lefloch} \&
  {Lazareff}}{1994}]{1994A&A...289..559L}
{Lefloch} B.,  {Lazareff} B.,  1994, \aap, 289, 559

\bibitem[\protect\citeauthoryear{{LeVeque}}{{LeVeque}}{2002}]{Leveque:2002}
{LeVeque} R.~J.,  2002, {Finite volume methods for hyperbolic problems }.
{Cambridge Texts in Applied Mathematics}, {Cambridge University Press}

\bibitem[\protect\citeauthoryear{{L{\'o}pez-Mart{\'{\i}}n}, {Raga}, {Mellema},
  {Henney} \& {Cant{\'o}}}{{L{\'o}pez-Mart{\'{\i}}n}
  et~al.}{2001}]{2001ApJ...548..288L}
{L{\'o}pez-Mart{\'{\i}}n} L.,  {Raga} A.~C.,  {Mellema} G.,  {Henney} W.~J.,
  {Cant{\'o}} J.,  2001, \apj, 548, 288

\bibitem[\protect\citeauthoryear{{Mac Low}, {Toraskar}, {Oishi} \& {Abel}}{{Mac
  Low} et~al.}{2007}]{2007ApJ...668..980M}
{Mac Low} M.-M.,  {Toraskar} J.,  {Oishi} J.~S.,    {Abel} T.,  2007, \apj,
  668, 980

\bibitem[\protect\citeauthoryear{{Mellema}, {Arthur}, {Henney}, {Iliev} \&
  {Shapiro}}{{Mellema} et~al.}{2006a}]{2006ApJ...647..397M}
{Mellema} G.,  {Arthur} S.~J.,  {Henney} W.~J.,  {Iliev} I.~T.,    {Shapiro}
  P.~R.,  2006a, \apj, 647, 397

\bibitem[\protect\citeauthoryear{{Mellema}, {Iliev}, {Alvarez} \&
  {Shapiro}}{{Mellema} et~al.}{2006b}]{2006NewA...11..374M}
{Mellema} G.,  {Iliev} I.~T.,  {Alvarez} M.~A.,    {Shapiro} P.~R.,  2006b, New
  Astronomy, 11, 374

\bibitem[\protect\citeauthoryear{{Mellema}, {Raga}, {Canto}, {Lundqvist},
  {Balick}, {Steffen} \& {Noriega-Crespo}}{{Mellema}
  et~al.}{1998}]{1998A&A...331..335M}
{Mellema} G.,  {Raga} A.~C.,  {Canto} J.,  {Lundqvist} P.,  {Balick} B.,
  {Steffen} W.,    {Noriega-Crespo} A.,  1998, \aap, 331, 335

\bibitem[\protect\citeauthoryear{{Minkowski}}{{Minkowski}}{1949}]{1949PASP...6%
1..151M}
{Minkowski} R.,  1949, \pasp, 61, 151

\bibitem[\protect\citeauthoryear{{Mizuta}, {Kane}, {Pound}, {Remington},
  {Ryutov} \& {Takabe}}{{Mizuta} et~al.}{2006}]{2006ApJ...647.1151M}
{Mizuta} A.,  {Kane} J.~O.,  {Pound} M.~W.,  {Remington} B.~A.,  {Ryutov}
  D.~D.,    {Takabe} H.,  2006, \apj, 647, 1151

\bibitem[\protect\citeauthoryear{{Motoyama}, {Umemoto} \& {Shang}}{{Motoyama}
  et~al.}{2007}]{2007A&A...467..657M}
{Motoyama} K.,  {Umemoto} T.,    {Shang} H.,  2007, \aap, 467, 657

\bibitem[\protect\citeauthoryear{{O'Dell}}{{O'Dell}}{2001}]{2001ARA&A..39...99%
O}
{O'Dell} C.~R.,  2001, \araa, 39, 99

\bibitem[\protect\citeauthoryear{{O'Dell}, {Balick}, {Hajian}, {Henney} \&
  {Burkert}}{{O'Dell} et~al.}{2002}]{2002AJ....123.3329O}
{O'Dell} C.~R.,  {Balick} B.,  {Hajian} A.~R.,  {Henney} W.~J.,    {Burkert}
  A.,  2002, \aj, 123, 3329

\bibitem[\protect\citeauthoryear{{O'Dell}, {Henney} \& {Ferland}}{{O'Dell}
  et~al.}{2005}]{2005AJ....130..172O}
{O'Dell} C.~R.,  {Henney} W.~J.,    {Ferland} G.~J.,  2005, \aj, 130, 172

\bibitem[\protect\citeauthoryear{{O'Dell} \& {Yusef-Zadeh}}{{O'Dell} \&
  {Yusef-Zadeh}}{2000}]{2000AJ....120..382O}
{O'Dell} C.~R.,  {Yusef-Zadeh} F.,  2000, \aj, 120, 382

\bibitem[\protect\citeauthoryear{{Oort} \& {Spitzer}}{{Oort} \&
  {Spitzer}}{1955}]{1955ApJ...121....6O}
{Oort} J.~H.,  {Spitzer} L.~J.,  1955, \apj, 121, 6

\bibitem[\protect\citeauthoryear{{Osterbrock}}{{Osterbrock}}{1957}]{1957ApJ...%
125..622O}
{Osterbrock} D.~E.,  1957, \apj, 125, 622

\bibitem[\protect\citeauthoryear{{Osterbrock} \& {Ferland}}{{Osterbrock} \&
  {Ferland}}{2006}]{2006agna.book.....O}
{Osterbrock} D.~E.,  {Ferland} G.~J.,  2006, {Astrophysics of gaseous nebulae
  and active galactic nuclei}, second edn.
Sausalito, CA: University Science Books

\bibitem[\protect\citeauthoryear{{Pavlakis}, {Williams}, {Dyson}, {Falle} \&
  {Hartquist}}{{Pavlakis} et~al.}{2001}]{2001A&A...369..263P}
{Pavlakis} K.~G.,  {Williams} R.~J.~R.,  {Dyson} J.~E.,  {Falle} S.~A.~E.~G.,
   {Hartquist} T.~W.,  2001, \aap, 369, 263

\bibitem[\protect\citeauthoryear{{Pottasch}}{{Pottasch}}{1956}]{1956BAN....13.%
..77P}
{Pottasch} S.~R.,  1956, \bain, 13, 77

\bibitem[\protect\citeauthoryear{{Raga}, {Henney}, {Vasconcelos}, {Cerqueira},
  {Esquivel} \& {Rodr{\'{\i}}guez-Gonz{\'a}lez}}{{Raga}
  et~al.}{2009}]{Raga:2009}
{Raga} A.~C.,  {Henney} W.,  {Vasconcelos} J.,  {Cerqueira} A.,  {Esquivel} A.,
     {Rodr{\'{\i}}guez-Gonz{\'a}lez} A.,  2009, \mnras, 392, 964

\bibitem[\protect\citeauthoryear{{Rathborne}, {Brooks}, {Burton}, {Cohen} \&
  {Bontemps}}{{Rathborne} et~al.}{2004}]{2004A&A...418..563R}
{Rathborne} J.~M.,  {Brooks} K.~J.,  {Burton} M.~G.,  {Cohen} M.,    {Bontemps}
  S.,  2004, \aap, 418, 563

\bibitem[\protect\citeauthoryear{{Redman}, {Williams}, {Dyson}, {Hartquist} \&
  {Fernandez}}{{Redman} et~al.}{1998}]{1998A&A...331.1099R}
{Redman} M.~P.,  {Williams} R.~J.~R.,  {Dyson} J.~E.,  {Hartquist} T.~W.,
  {Fernandez} B.~R.,  1998, \aap, 331, 1099

\bibitem[\protect\citeauthoryear{{Reipurth}}{{Reipurth}}{1983}]{1983A&A...117.%
.183R}
{Reipurth} B.,  1983, \aap, 117, 183

\bibitem[\protect\citeauthoryear{{Ryutov}, {Kane}, {Mizuta}, {Pound} \&
  {Remington}}{{Ryutov} et~al.}{2005}]{2005Ap&SS.298..183R}
{Ryutov} D.~D.,  {Kane} J.~O.,  {Mizuta} A.,  {Pound} M.~W.,    {Remington}
  B.~A.,  2005, \apss, 298, 183

\bibitem[\protect\citeauthoryear{{Sandford} II, {Whitaker} \&
  {Klein}}{{Sandford} et~al.}{1982}]{1982ApJ...260..183S}
{Sandford} II M.~T.,  {Whitaker} R.~W.,    {Klein} R.~I.,  1982, \apj, 260, 183

\bibitem[\protect\citeauthoryear{{Shaw}, {Ferland}, {Abel}, {Stancil} \& {van
  Hoof}}{{Shaw} et~al.}{2005}]{Shaw:2005}
{Shaw} G.,  {Ferland} G.~J.,  {Abel} N.~P.,  {Stancil} P.~C.,    {van Hoof}
  P.~A.~M.,  2005, \apj, 624, 794

\bibitem[\protect\citeauthoryear{{Spitzer}}{{Spitzer}}{1954}]{1954ApJ...120...%
.1S}
{Spitzer} L.~J.,  1954, \apj, 120, 1

\bibitem[\protect\citeauthoryear{{Sridharan}, {Bhatt} \&
  {Rajagopal}}{{Sridharan} et~al.}{1996}]{1996MNRAS.279.1191S}
{Sridharan} T.~K.,  {Bhatt} H.~C.,    {Rajagopal} J.,  1996, \mnras, 279, 1191

\bibitem[\protect\citeauthoryear{{Stelzer}, {Flaccomio}, {Montmerle}, {Micela},
  {Sciortino}, {Favata}, {Preibisch} \& {Feigelson}}{{Stelzer}
  et~al.}{2005}]{2005ApJS..160..557S}
{Stelzer} B.,  {Flaccomio} E.,  {Montmerle} T.,  {Micela} G.,  {Sciortino} S.,
  {Favata} F.,  {Preibisch} T.,    {Feigelson} E.~D.,  2005, \apjs, 160, 557

\bibitem[\protect\citeauthoryear{{Stoerzer} \& {Hollenbach}}{{Stoerzer} \&
  {Hollenbach}}{1998}]{1998ApJ...495..853S}
{Stoerzer} H.,  {Hollenbach} D.,  1998, \apj, 495, 853

\bibitem[\protect\citeauthoryear{{Str{\"o}mgren}}{{Str{\"o}mgren}}{1939}]{1939%
ApJ....89..526S}
{Str{\"o}mgren} B.,  1939, \apj, 89, 526

\bibitem[\protect\citeauthoryear{{Tenorio-Tagle} \& {Bedijn}}{{Tenorio-Tagle}
  \& {Bedijn}}{1981}]{1981A&A....99..305T}
{Tenorio-Tagle} G.,  {Bedijn} P.~J.,  1981, \aap, 99, 305

\bibitem[\protect\citeauthoryear{{T{\'o}th}}{{T{\'o}th}}{2000}]{2000JCoPh.161.%
.605T}
{T{\'o}th} G.,  2000, Journal of Computational Physics, 161, 605

\bibitem[\protect\citeauthoryear{{Vall{\'e}e} \& {Fiege}}{{Vall{\'e}e} \&
  {Fiege}}{2007}]{2007AJ....134..628V}
{Vall{\'e}e} J.~P.,  {Fiege} J.~D.,  2007, \aj, 134, 628

\bibitem[\protect\citeauthoryear{{Vall{\'e}e}, {Greaves} \&
  {Fiege}}{{Vall{\'e}e} et~al.}{2003}]{2003ApJ...588..910V}
{Vall{\'e}e} J.~P.,  {Greaves} J.~S.,    {Fiege} J.~D.,  2003, \apj, 588, 910

\bibitem[\protect\citeauthoryear{{van Hoof}, {Weingartner}, {Martin}, {Volk} \&
  {Ferland}}{{van Hoof} et~al.}{2004}]{2004MNRAS.350.1330V}
{van Hoof} P.~A.~M.,  {Weingartner} J.~C.,  {Martin} P.~G.,  {Volk} K.,
  {Ferland} G.~J.,  2004, \mnras, 350, 1330

\bibitem[\protect\citeauthoryear{{V{\'a}zquez-Semadeni}, {G{\'o}mez},
  {Jappsen}, {Ballesteros-Paredes}, {Gonz{\'a}lez} \&
  {Klessen}}{{V{\'a}zquez-Semadeni} et~al.}{2007}]{2007ApJ...657..870V}
{V{\'a}zquez-Semadeni} E.,  {G{\'o}mez} G.~C.,  {Jappsen} A.~K.,
  {Ballesteros-Paredes} J.,  {Gonz{\'a}lez} R.~F.,    {Klessen} R.~S.,  2007,
  \apj, 657, 870

\bibitem[\protect\citeauthoryear{{Whalen} \& {Norman}}{{Whalen} \&
  {Norman}}{2008}]{2008ApJ...672..287W}
{Whalen} D.~J.,  {Norman} M.~L.,  2008, \apj, 672, 287

\bibitem[\protect\citeauthoryear{{Williams}, {Bergin}, {Caselli}, {Myers} \&
  {Plume}}{{Williams} et~al.}{1998}]{Williams:1998}
{Williams} J.~P.,  {Bergin} E.~A.,  {Caselli} P.,  {Myers} P.~C.,    {Plume}
  R.,  1998, \apj, 503, 689

\bibitem[\protect\citeauthoryear{{Williams}}{{Williams}}{1999}]{1999MNRAS.310.%
.789W}
{Williams} R.~J.~R.,  1999, \mnras, 310, 789

\bibitem[\protect\citeauthoryear{{Williams}}{{Williams}}{2002}]{2002MNRAS.331.%
.693W}
{Williams} R.~J.~R.,  2002, \mnras, 331, 693

\bibitem[\protect\citeauthoryear{{Williams}}{{Williams}}{2007}]{2007Ap&SS.307.%
.179W}
{Williams} R.~J.~R.,  2007, \apss, 307, 179

\bibitem[\protect\citeauthoryear{{Williams} \& {Dyson}}{{Williams} \&
  {Dyson}}{2001}]{2001MNRAS.325..293W}
{Williams} R.~J.~R.,  {Dyson} J.~E.,  2001, \mnras, 325, 293

\bibitem[\protect\citeauthoryear{{Williams}, {Dyson} \& {Hartquist}}{{Williams}
  et~al.}{2000}]{2000MNRAS.314..315W}
{Williams} R.~J.~R.,  {Dyson} J.~E.,    {Hartquist} T.~W.,  2000, \mnras, 314,
  315

\bibitem[\protect\citeauthoryear{{Williams}, {Ward-Thompson} \&
  {Whitworth}}{{Williams} et~al.}{2001}]{2001MNRAS.327..788W}
{Williams} R.~J.~R.,  {Ward-Thompson} D.,    {Whitworth} A.~P.,  2001, \mnras,
  327, 788

\bibitem[\protect\citeauthoryear{{Wolf}, {Launhardt} \& {Henning}}{{Wolf}
  et~al.}{2003}]{2003ApJ...592..233W}
{Wolf} S.,  {Launhardt} R.,    {Henning} T.,  2003, \apj, 592, 233

\end{thebibliography}

\appendix

\New{
\section{Treatment of gas heating and cooling in the Phab-C\textsuperscript{2} code}
\label{sec:heating-cooling-laws}

\begin{figure*}
  \centering
  \includegraphics[width=\linewidth]{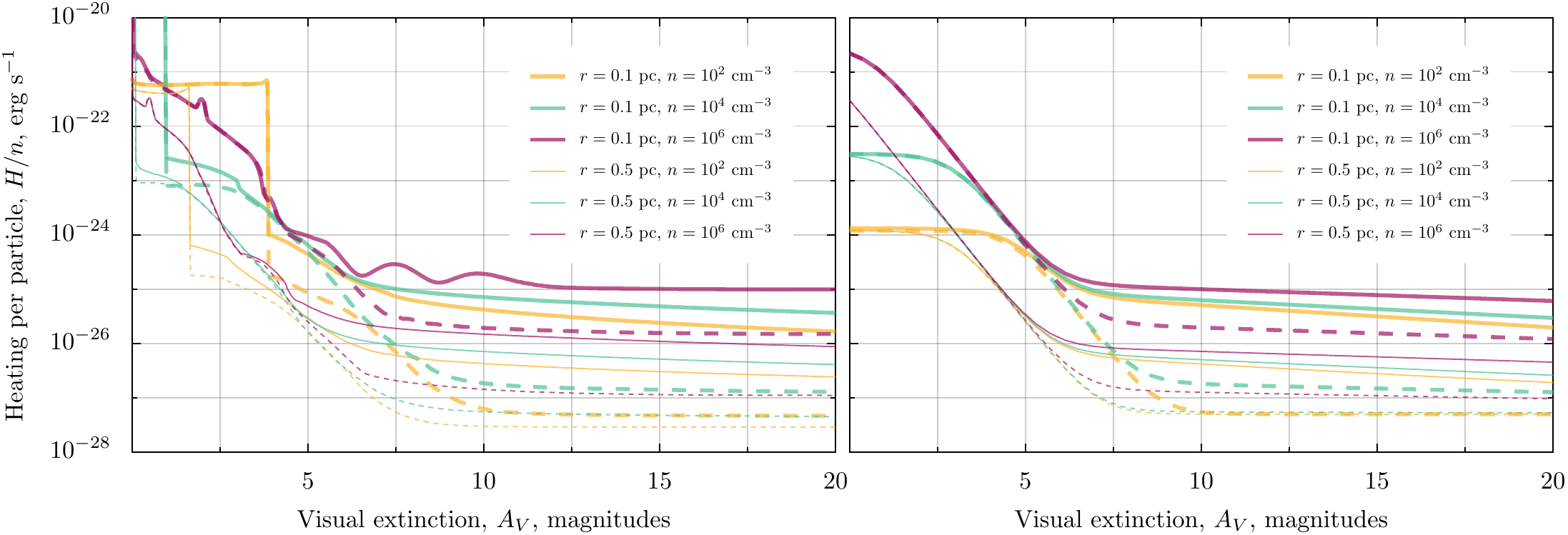}
 \caption[]{Heating versus \(A_V\) for Cloudy models (left panel) and simplified fits (right panel). Solid lines show models including x-ray illumination, dashed lines show models with no x~rays. Other parameters of the Cloudy models are shown in the key, see text for further details.}
  \label{fig:heat}
\end{figure*}
\begin{figure*}
  \centering
  \includegraphics[width=\linewidth]{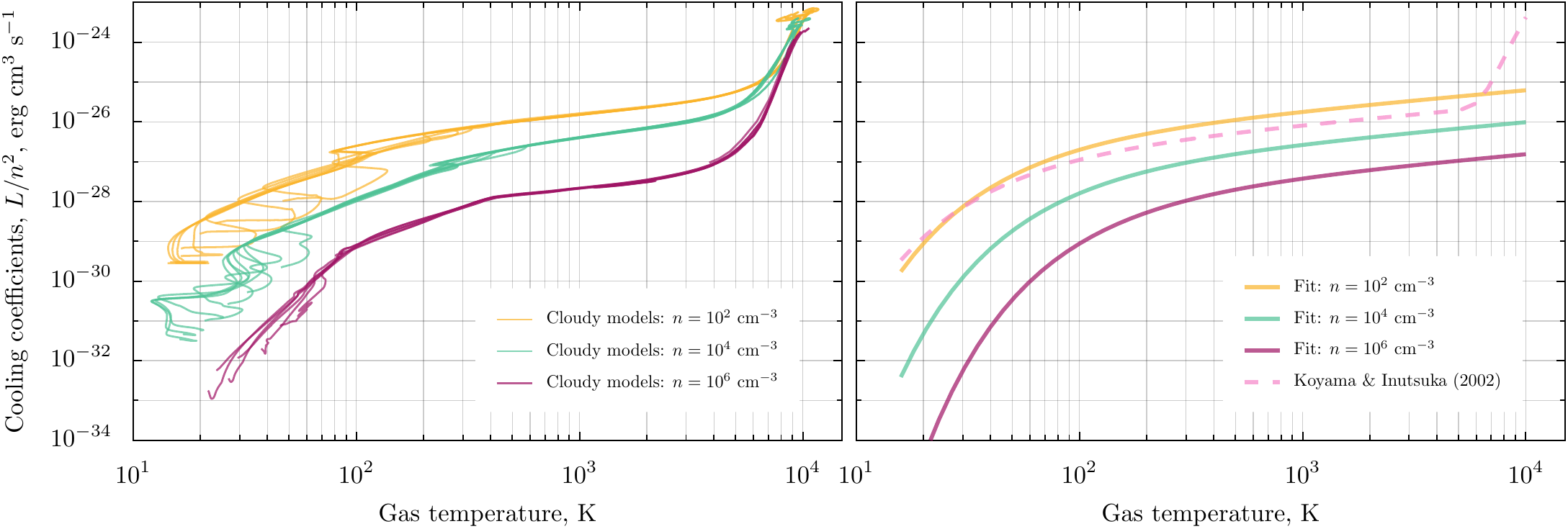}
 \caption[]{Cooling versus \(T\) from Cloudy models (left panel), together with simplified fits (right panel).}
  \label{fig:cool}
\end{figure*}

\subsection{Heating}
\label{sec:heating}

The volumetric heating of the gas in our numerical simulations is the sum of various terms, each of which represents the effects of a different type of radiation:
\begin{equation}
  \label{eq:heating-terms}
  H =  H_{\mathrm{EUV}} + H_{\mathrm{FUV}} + H_{\mathrm{X}} + H_{\mathrm{IR}}  + H_{\mathrm{CR}}
\end{equation}

For the gas heating by ionising EUV photons \(H_{\mathrm{EUV}}\), we consider only the photoionisation heating of hydrogen, which is calculated in a similar way to the photoionisation rate in equation~(\ref{mhd5}) of \S~\ref{sec:basic-equations}: 
\begin{equation}
  \label{eq:photoelectric}
  H_{\mathrm{EUV}} = n\Hz \int_{\nu_0}^\infty \!\!\!\sigma_\nu (4 \pi J_\nu / h\nu) (h\nu - h\nu_0) \,d\nu  \quad \heatunits{}
\end{equation}
where \(n\Hz\) is the number density of neutral hydrogen, \(\sigma_\nu\) is the photoionisation cross-section and \(J_\nu\) is the local mean intensity of the ionising radiation field, both functions of the photon frequency \(\nu\), as discussed in \S~\ref{sec:basic-equations}. Note that the hardening of the radiation field as one approaches the ionisation front is fully taken into consideration. In fully ionised gas, \(H_{\mathrm{EUV}}\) dominates the total heating by many orders of magnitude. 

In neutral and molecular gas, the ionising radiation is no longer present and other heating mechanisms become important. All the remaining terms that contribute to the total heating were determined by running a large grid of static, constant-density photodissociation-region (PDR) models using the plasma physics code Cloudy \citep{1998PASP..110..761F}, in which the gas density and incident radiation field were varied over a wide range (see Fig.~\ref{fig:heat} and \S~\ref{sec:validation-fits}). In all these models, the gas-phase abundances and dust properties were set at those appropriate to the Orion Nebula. 

For gas close behind the ionisation front, non-ionising far-ultraviolet (FUV) radiation with photon energies of around 6 to 13.6~eV is the most important heating agent, principally via absorption by dust grains (including PAHs) and molecular hydrogen lines. We find that the resultant heating rate, \(H_{\mathrm{FUV}}\), can be approximated as 
% DONE 11 Mar 2009 \TODO{(still need to check numerical factor in denominator for consistency with code)}
\begin{equation}
  \label{eq:fuv}
  H_{\mathrm{FUV}} = \frac{1.9 \times 10^{-26} n G_0 e^{-1.9 A_V}} { 1 + 6.4 (G_0/n) e^{-1.9 A_V} }  \quad \heatunits
\end{equation}
where \(G_0\) is the strength of the FUV radiation field in units of the Habing flux (\(1.2\times 10^7~\mathrm{photons\ cm^{-2}\ s^{-1}}\) in the range 912 to 2000~\AA{}) and \(A_V\) is the visual-band optical extinction in magnitudes: \(A_V = 1.086 \sigma_V N\), in which \(\sigma_V\) is the dust extinction cross-section (assumed to be \(5\times 10^{-22}~\mathrm{cm^2\ H^{-1}}\) \citealp{1991ApJ...374..580B}) and \( N = \int n d\ell\) is the column density of hydrogen nucleons.  When the gas density, \(n\) in \(\pcc\), is large compared with the local FUV field, \(G_0 e^{-1.9 A_V}\), the denominator in equation~(\ref{eq:fuv}) is close to unity, in which case the per-particle heating rate, \(H_{\mathrm{FUV}}/n\) is proportional to the local FUV intensity.\footnote{\New{Note that a different grain size distribution from the Orion-type grains used here would give a steeper decline with \(A_V\).}} For smaller densities than this, the gas-grain thermal coupling is inefficient, leading to a reduction in the heating rate. FUV heating makes an important contribution in PDRs to a depth of \(A_V = 2\)--\(6\) depending on the strength of the incident radiation field. 

For a single point source of radiation,the unextinguished FUV field is calculated in a similar way as for the ionising radiation in eq.~(\ref{eq:rad}):
\begin{equation}
  \label{eq:fuvfield}
  1.2\times 10^7 G_0 = \frac{Q_{\mathrm{FUV}}}{4 \pi |\vec r - \vec r_*|^2} , 
\end{equation}
where \(Q_{\mathrm{FUV}}\) is the FUV photon luminosity of the source. For a typical O~star, \(Q_{\mathrm{FUV}} / Q_{\mathrm{H}}\) is in the range 0.5 to 1.0, depending on the spectral type. The column density, \(N\), which is required in order to determine the local extinction, \(A_V\), is calculated by a short-characteristic method in the same way as for the ionising radiation \citep{2006NewA...11..374M}, but using the total hydrogen density in place of the neutral hydrogen density. In this treatment, the diffuse non-ionising radiation is simply ignored, although it could in principle be accounted for by replacing the term \(G_0 e^{-1.9 A_V}\) in equation~(\ref{eq:fuv}) by a local intensity calculated from a more realistic solution of the radiative transfer. 

Hard x~rays easily penetrate large columns of gas and can therefore be important heating agents deep inside the PDR, in fully molecular gas. We have carried out fits to Cloudy models illuminated by x~rays from collisional ionisation equilibrium plasmas with temperatures between \(6 \times 10^6\) and \(4 \times 10^7\)~K, determining that an appropriate approximate heating rate is
\begin{equation}
  \label{eq:xray-heating}
  H_{\mathrm{X}} = 6 \times 10^{-23} n F_{\mathrm{X}} \quad \heatunits , 
\end{equation}
where \(F_{\mathrm{X}}\) is the unattenuated x-ray flux, with units \(\mathrm{erg\ cm^{-2}\ s^{-1}}\), in the band from 0.5 to 8.0~keV\@. The Cloudy models extend to \(A_V = 20\) and show no clear fall-off of \(H_{\mathrm{X}}\) with \(A_V\), except for at low gas densities. However, we caution against applying equation~(\ref{eq:xray-heating}) at significantly higher columns than this, since attenuation of the x~rays must become important eventually. 

For sufficiently high gas densities (\(n \ge 10^4~\pcc\)) the gas and grains are well coupled, so that stellar radiation that is re-processed to the infrared at \(A_V \sim 1\) and then re-absorbed by dust at larger columns will indirectly heat the gas there. From fits to Cloudy models, we determine a heating rate for this component of
\begin{equation}
  \label{eq:irheat}
  H_{\mathrm{IR}} = 7.7 \times 10^{-32} n (1+ n_1/n)^{-2} G_0 e^{-0.05 A_V} \quad \heatunits , 
\end{equation}
where \(n_1 = 3\times 10^4~\pcc\). 

Finally, when the radiation field is very weak, the baseline heating is given by cosmic rays, which we assume to give a constant per-particle heating rate such that
\begin{equation}
  \label{eq:cosmicrayheating}
  H_{\mathrm{CR}} = 5 \times 10^{-28} n \quad \heatunits . 
\end{equation}

\subsection{Cooling}
\label{sec:cooling}
The radiative cooling of the gas is also a sum of various different terms: 
\begin{equation}
    \label{eq:cooling}
    L = L_{\mathrm{M}^+} +  L_{\mathrm{M}^0}+ L_{\mathrm{H}^+}  + L_{\mathrm{H}^0}  + L_{\mathrm{CIE}} + L_{\mathrm{PDR}}
  \end{equation}
In fully ionised gas, collisionally excited optical lines of ionised metals dominate the cooling. We approximately model this cooling term, \(L_{\mathrm{M}^+}\) by multiplying the emission of the typically dominant line ([\ion{O}{2}]~4363~\AA{}) by a factor 3 to account for all the other lines \citep[Appendix~A]{Biro:1995}.
\begin{equation}
  \label{eq:ionized-metal-cool}
   L_{\mathrm{M}^+} =  2.905\times 10^{-19} z_{\mathrm{O}} n_{\mathrm{e}} n\Hp e^{ -T_1/T - (T_2/T)^2} \quad\heatunits,
\end{equation}
where \(z_{\mathrm{O}}\) is the oxygen abundance, \(T_1 = 33,610\)~K and \(T_2 = 2180\)~K\@. We ignore any changes in the cooling rate due to second and higher ionization of metals (for instance, from O\(^+\) to O\(^{++}\) in the interior of the ionized region, simply assuming that the strongest cooling lines always follow equation~(\ref{eq:ionized-metal-cool}). A similar approach is used for collisionally excited lines of neutral metals:
\begin{equation}
  \label{eq:neutral-metal-cool}
  L_{\mathrm{M}^0} =  4.477\times 10^{-20} z_{\mathrm{O}} n_{\mathrm{e}} n\Hz e^{ -T_3/T + (T_4/T)^2} \quad\heatunits, 
\end{equation}
where \(T_3 = 28,390\)~K and \(T_4 = 1780\)~K\@. The product \(n_{\mathrm{e}} n\Hz\) is proportional to \( x_{\mathrm{i}} (1 - x_{\mathrm{i}}) \), where \(x_{\mathrm{i}}\) is the hydrogen ionization fraction, so that \(L_{\mathrm{M}^0}\) only makes a significant contribution inside ionization or recombination fronts, where \(x_{\mathrm{i}} \sim 0.5\). 

We also include cooling terms due to free-free and free-bound transitions of ionized hydrogen 
\begin{equation}
  \label{eq:ionized-H-cool}
   L_{\mathrm{H}^+} = n_{\mathrm{e}} n\Hp f_{\mathrm{H}^+}(T) \quad \heatunits, 
\end{equation}
and collisionally excited lines of neutral hydrogen
\begin{equation}
  \label{eq:neutral-H-cool}
   L_{\mathrm{H}^0} = n_{\mathrm{e}} n\Hz f_{\mathrm{H}^0}(T) \quad \heatunits ,  
\end{equation}
where \(f_{\mathrm{H}^+}(T)\) and \(f_{\mathrm{H}^0}(T)\) are taken from \citet{1994MNRAS.268..109H}. 

For high gas temperatures (above \(50,000~\mathrm{K}\)) we adopt the following fit to the collisional ionisation equilibrium cooling curve: 
\begin{equation}
  \label{eq:cie-cool}
  L_{\mathrm{CIE}} = 3.485 \times 10^{-15} z_{\mathrm{O}} T^{-0.63} ( 1 - e^{-(10^{-5} T)^{1.63}})  \quad \heatunits, 
\end{equation}
which is mainly due to collisional lines of highly ionized metals (e.g., \citealp{1972ARA&A..10..375D}). Note that this term does not include the bremsstrahlung contribution that becomes important at very high temperatures, since that is already included in \(L_{\mathrm{H}^+}\). 

For neutral and molecular gas, we adopt a similar approach to that described above for the heating, fitting the cooling determined from Cloudy PDR models as a function of gas density and temperature (Fig.~\ref{fig:cool}). The resultant fit is
\begin{equation}
  \label{eq:pdr-cool}
  L_{\mathrm{PDR}} = 3.981\times 10^{-27} n^{1.6} T^{0.5} e^{-T_0(n)/T} \quad \heatunits, 
\end{equation}
where \(T_0(n)  = 70 + 220 (n/10^6)^{0.2}\). 
}

\New{

\subsection{Validation of simplified fits to equilibrium PDR heating and cooling}
\label{sec:validation-fits}

Figure~\ref{fig:heat} compares the simplified heating functions that we have adopted for neutral/molecular gas, \(H_{\mathrm{FUV}} + H_{\mathrm{X}} + H_{\mathrm{IR}}  + H_{\mathrm{CR}}\), with the results of detailed PDR models calculated with Cloudy. One set of Cloudy models was calculated using the following components for the incident radiation: 
\begin{enumerate} 
\item\label{item:photosphere} A 40,000~K black body with \(Q_\mathrm{H} = 10^{49}~\mathrm{s}^{-1}\) to represent the O~star photosphere.
\item\label{item:stellarX} An optically thin x-ray source with luminosity \(L_\mathrm{X,1} = 2.29 \times 10^{33}~\mathrm{erg\ s^{-1}}\) and temperature \(T_\mathrm{X,1} = 1.6 \times 10^7~\mathrm{K}\) to represent circumstellar emission from the O~star \citep{2005ApJS..160..379F, 2005ApJS..160..557S}.
\item\label{item:clusterX} A second optically thin x-ray source with luminosity \(L_\mathrm{X,2} = 3.47 \times 10^{33}~\mathrm{erg\ s^{-1}}\) and \(T_\mathrm{X,2} = 2.5\times 10^7~\mathrm{K}\) to represent chromospheric emission from low-mass stars in the accompanying star cluster \citep{2003ApJ...582..382F}.
\item\label{item:CR} The standard cosmic ray background, equivalent to an H\(_2\) ionization rate of \(5 \times 10^{-17}~\mathrm{s^{-1}}\) \citep{Williams:1998}
% \item The local interstellar radiation field \citep{Black1987}. % Black, J. H. 1987, in Interstellar Processes, ed. D.J. Hollenbach & H.A. Thronson, (Dordrecht: Reidel), p 731 
\end{enumerate}
A second set of Cloudy models was calculated with a radiation field identical to the above but without any x~rays. In both cases, the gas phase abundances and grain properties are assumed to be those found in the Orion Nebula \citep{1991ApJ...374..580B, 2004MNRAS.350.1330V}, including a PAH component as in \citet{Abel:2008}. A detailed model of molecular hydrogen is included in the calculations, as described in \citep{Shaw:2005}. The models are all calculated for a plane-parallel geometry, with the total hydrogen density held constant at \(10^2\), \(10^4\), or \(10^6~\pcc\), with the incident flux calculated for distances of \(r = 0.1\), \(0.2\), \(0.5\), and \(1.0\)~pc from the ionizing star. For components \ref{item:photosphere} and \ref{item:stellarX}, the flux is calculated assuming a point source of radiation, whereas for component \ref{item:clusterX}, the source is assumed to be extended over the stellar cluster or radius \(r_{\mathrm{c}} = 0.3~\mathrm{pc}\) and the flux is approximated as \(F_{\mathrm{X,2}} = L_{\mathrm{X,2}} / 4 \pi (r_{\mathrm{c}}^2 + r^2)\). 

Figure~\ref{fig:heat} shows that our approximate heating functions do a very good job of reproducing the results of the Cloudy models. Note that the lower density Cloudy models include an extended fully ionized portion at low \(A_V\) (visible as a plateau of high constant heating in the graph), in which the heating will be dominated by \(H_{\mathrm{EUV}}\), which is not included in the fits shown in the right panel. Apart from in the ionized gas, the region with \(A_V < 6\)--\(10\) is dominated by \(H_{\mathrm{FUV}}\), whereas \(H_{\mathrm{X}}\) dominates at higher depths for all models that include x~rays. In the models without x~rays, it is \(H_{\mathrm{IR}}\) that dominates at depth, except for at low densities or large distances from the star, where \(H_{\mathrm{CR}}\) becomes important. Additional features are seen in the heating curves of the Cloudy models that correspond to the dissociation fronts of molecules, particularly H\(_2\) and CO\@. No attempt is made to reproduce these low-amplitude features in our fits. Another caveat is that the Cloudy models that we have used in constructing our fits all assume static equilibrium and do not take into account advective and time-dependent effects. Such effects will be most important at discontinuities, where they tend to increase the heating rate with respect to the equilibrium calculation \citep{2005ApJ...621..328H}. In the case of the ionization front, this effect is fully taken into account in our simulations since the neutral hydrogen density that enters into \(H_{\mathrm{EUV}}\) (eq.~(\ref{eq:photoelectric})) is calculated from the fully time-dependent equation~(\ref{mhd5}). The molecular hydrogen dissociation front, on the other hand, is not explicitly included in our fits and advective effects  there are not accounted for. However, for the case of photodissociation regions illuminated by O~stars, the changes introduced by advection are expected to be rather small \citep{1998ApJ...495..853S}. A notable exception to this will occur in cases where the ionization front and dissociation front merge, which tends to occur for low ionization parameters and hard incident spectra \citep{1996ApJ...458..222B}. In such cases, a much more careful treatment of the heating is necessary, as in \citet{2007ApJ...671L.137H}.

Figure~\ref{fig:cool} presents a similar comparison for our approximate cooling function \(L_{\mathrm{PDR}}\). It can be seen that this function does a good job of reproducing the principal variation of the cooling with density and temperature. At temperatures below \(\sim 100\)~K, the cooling is prinicipally due to millimetre CO lines (at higher densities) or far-infrared neutral C lines (at lower densities), while at warmer temperatures it is dominated by far-infrared neutral O lines (at higher densities) or singly ionised C lines (at lower densities). At temperatures greater that \(3000\)~K, the cooling begins to be dominated by the optical and UV lines that we include in \(L_{\mathrm{H}^0}\) and \(L_{\mathrm{M}^0}\) using the non-equilibrium time-dependent ionization fractions, so this portion of the cooling is not included in the equilibrium fits shown in the right panel. At the lowest temperatures, one sees that the cooling in the Cloudy models is no longer uniquely determined by \(T\) and \(n\), especially for the lower densities. Most of this extra variation, which we make no attempt to reproduce in our fits, is due to changes in the molecular abundances. Also shown in the figure is the curve of \citet{2002ApJ...564L..97K}, as corrected by \citet{2007ApJ...657..870V}, which has been frequently employed in numerical simulations:
\begin{multline}
  \label{eq:KI}
  L_{\mathrm{KI}}(n, T) = 2 \times 10^{-19} n^2 \left(e^{-1.184 \times 10^5/(T+1000)} \right. \\ 
  \left. {} + 1.4\times 10^{-9} T^{1/2} e^{-92/T}\right) \quad \heatunits
\end{multline}
It can be seen that this curve is a reasonable approximation to the cooling at low densities, but that it seriously overestimates the cooling rate for \(n > 10^3~\pcc\), being 100 times too large for \(T = 100\)~K, \(n = 10^6~\pcc\). 

In summary, we believe that the heating and cooling functions that we have introduced in this appendix represent a substantial improvement over what has been commonly used in the previous literature, which has generally either assumed an isothermal equation of state for neutral/molecular  gas (e.g., \citealp{2007MNRAS.377..383E}) or used \(H = H_{\mathrm{CR}}\) and \(L = L_{\mathrm{KI}}\) (e.g., \citealp{2007ApJ...671..518K}). The typical neglect of FUV/optical dust heating (important for columns with \(A_V < 5\)),  of x-ray heating (important within \(\simeq 3\)~pc of a typical O~star) and of collisional deexcitation of the principal cooling lines (important for \(n > 10^4~\pcc\)) will tend to result in an underestimate of the temperature of shocked neutral/molecular gas. Since we have broken down the heating function into terms due to different wavelength bands, our results can be used in the case of illumination by stars of different effective temperatures and with different x-ray luminosities.

}

\bsp

\end{document}